# Rapid Binary Mass Transfer: Circumbinary Outflows and Angular Momentum Losses

Peter Scherbak[1], Wenbin Lu[2], and Jim Fuller[3]
[1] California Institute of Technology, Astronomy Department, Pasadena, CA 91125, USA
[2] Departments of Astronomy and Theoretical Astrophysics Center, UC Berkeley, Berkeley, CA 94720, USA
[3] TAPIR, California Institute of Technology, Pasadena, CA 91125, USA

## Abstract

High rates of stable mass transfer likely occur for some binary star systems, but the resulting flow of mass and angular momentum (AM) is unclear. We perform hydrodynamical simulations of a polytropic donor star and a point-mass secondary to determine the mass, AM, and velocity of gas that leaves the system, and the dependence on binary parameters such as mass ratio. The simulations use an adiabatic equation of state and do not include radiative cooling or irradiation of the outflow. Mass transfer is initiated by injecting heat into the stellar envelope, causing it to gradually inflate and overflow its Roche lobe. The transferred mass flows into an accretion disk, but soon begins to escape through the outer Lagrange point (L2), with a lesser amount escaping through the L3 point. This creates an equatorially concentrated circumbinary outflow with an opening angle of 10°–30° with a wind-like density profile $\rho \propto r^{-2}$. We find that the ratios of the specific AM of the outflowing gas over that of the L2 point are approximately {0.95, 0.9, 0.8, 0.65} for mass ratios $q =$ {0.25, 0.5, 1, 2} (accretor/donor). The asymptotic radial velocity of the outflowing gas, in units of the binary orbital velocity, is approximately 0.1–0.2 for the same mass ratios, except for $q = 0.25$, where it might be higher. This outflow, if ultimately unbound from the binary, may be a source of circumstellar material that interacts with ejecta from a subsequent supernova or stellar merger.

## 1. Introduction

Mass transfer (MT) occurring between a donor star and a secondary star is common in binary star systems, especially for binaries containing massive stars (H. Sana et al. 2012). Episodes of MT precede the formation of isolated compact binaries (K. A. Postnov & L. R. Yungelson 2014), including gravitational-wave sources currently being discovered by LIGO (LIGO Scientific Collaboration et al. 2023) and sources that will be discovered by LISA (V. Korol et al. 2022). In addition, a large portion of core-collapse supernovae occur in binaries that underwent MT in the past (F. R. N. Schneider et al. 2021).

There are multiple uncertainties regarding MT that make predicting the fate of a mass-transferring system difficult. The criteria for MT stability, and whether a common envelope phase occurs (B. Paczynski 1976; N. Ivanova et al. 2013), remains unclear. The rate of MT can be estimated based on prescriptions applied to 1D stellar evolution models (e.g., U. Kolb & H. Ritter 1990), but recent works appear to disagree with existing prescriptions (J. Cehula & O. Pejcha 2023; N. Ivanova et al. 2024). Another uncertainty is whether MT is conservative (S. S. Huang 1963; S. E. Mink et al. 2007), i.e., what fraction of the material is accreted by the secondary.

When nonconservative MT occurs, a major uncertainty is the angular momentum carried away by the escaping material. Nonconservative MT has been predicted to increase the fraction of binaries undergoing unstable MT and merging (R. Willcox et al. 2023). Many works assume that the specific angular momentum (angular momentum per unit mass lost) is equal to that of the orbit of the accreting star. However, if mass escapes by flowing through the outer Lagrange point, L2, then its specific angular momentum may be comparable to that of the L2 point (e.g., S. S. Huang 1963; M. MacLeod et al. 2018a), which is substantially larger than the specific angular momentum of the accretor. In the case of a massive binary with two stellar components, mass loss near L2 has been predicted as a likely outcome (P. Podsiadlowski et al. 1992), and the specific angular momentum of the outflow has been noted as an important uncertainty (S. Wellstein et al. 2001). M. Gallegos-Garcia et al. (2024; see also W. Lu et al. 2023) showed that, when the accretor is much more massive than the donor, the specific angular momentum lost through winds launched from the outer accretion disk can also be much larger than that of the accretor's orbit.

Several recent works have found that stable MT is an important channel to forming binary black holes (BBHs; K. Pavlovskii et al. 2017; M. Gallegos-Garcia et al. 2021; J. Klencki et al. 2021; P. Marchant et al. 2021; L. A. C. van Son et al. 2022; A. Picco et al. 2024). Of these, P. Marchant et al. (2021) and A. Picco et al. (2024) noted that L2 mass loss may precede BBH formation and that the amount of specific angular momentum lost is an outstanding uncertainty that impacts the predicted BBH population.

Recently, W. Lu et al. (2023) found that, for high rates of MT ($\gtrsim 10^{-4} M_\odot\,\mathrm{yr}^{-1}$), gas in the accretion disk cools inefficiently and is energetic enough such that it is likely to escape through L2 and form a circumbinary outflow. There are several classes of binaries that may achieve the high MT rates for which the assumptions of W. Lu et al. (2023) are valid, including: (1) binaries where the donor star crosses the Hertzsprung gap on a short timescale (M. S. Hjellming & R. F. Webbink 1987; P. Marchant et al. 2021; J. Klencki et al. 2022; K. D. Temmink et al. 2023; A. Dorozsmai & S. Toonen 2024); (2) He stars with a neutron star companion, which can initiate high rates of MT following core-He burning, known as case BB MT (J. D. M. Dewi et al. 2002; T. M. Tauris et al. 2015, 2017); and (3) a pre-supernova star in a binary, which expands a short time before the explosion due to late stages of nuclear burning or wave heating (E. Quataert & J. Shiode 2012; L. Mcley & N. Soker 2014; J. Fuller 2017; S. Wu & J. Fuller 2020; S. C. Wu & J. Fuller 2022). Observationally, L2 mass loss is likely occurring in binaries such as SS433 (K. M. Blundell et al. 2001; A. M. Cherepashchuk et al. 2018) and W Serpentis (K. Shepard et al. 2024).

If the donor star is a supernova progenitor, L2 mass loss resulting in a circumbinary outflow is a source of circumstellar material (CSM), which the supernova ejecta later interacts with. There exists a variety of observational evidence that a large fraction of core-collapse supernovae interact with CSM that occurred from pre-supernova mass loss, like narrow-line emission (e.g., F. Taddia et al. 2013) or faster rise times and more luminous emission at early times (e.g., P. Clark et al. 2020), but the mechanisms behind CSM formation remain unclear. It is important to quantify the speed and geometry of the circumbinary outflow to predict its observable effects on supernovae interacting with dense CSM.

In this paper, we extend the work of W. Lu et al. (2023) by performing hydrodynamical models of stable MT, to ascertain whether circumbinary outflows can form near L2. We also aim to determine their morphology, and to compute the angular momentum and the velocity of the ejected material. We use the PLUTO hydrodynamic code (A. Mignone et al. 2007, 2012) to examine MT between a donor star and a secondary, as well as any outflows from the accretion disk that result. The simulation domain includes a donor star, a point-mass accretor star, and an extended region about both stars.

There have been previous simulations that have involved outflows through L2, but with differences to our setup. O. Pejcha et al. (2016a, 2016b) performed smoothed particle hydrodynamic (SPH) simulations of L2 mass loss, and confirmed the prediction of F. H. Shu et al. (1979) regarding the binary mass ratios where material is bound or unbound. D. Hubová & O. Pejcha (2019) then broadened the initial conditions of their SPH simulations to include material with position offsets or velocity offsets from L2, finding a wider range of outcomes.

Previous simulations of mass transfer that have involved mass loss through L2 and sometimes L3 include J. L. A. Nandez et al. (2014), Z. Chen et al. (2017), M. MacLeod et al. (2018a, 2018b), K. Kadam et al. (2018), T. A. Reichardt et al. (2019), M. MacLeod & A. Loeb (2020a, 2020b), and D. Toyouchi et al. (2024). Several works investigating eccentric binaries and episodic MT also noted the occurrence of L2 mass loss (E. Regös et al. 2005; R. P. Church et al. 2009; C.-P. Lajoie & A. Sills 2010a), as did work focused on wind RLOF (S. Mohamed & P. Podsiadlowski 2012). In contrast to these works, we focus on the approximately steady-state behavior of stable MT where we track the outflow of gas from the binary.

This paper is organized as follows. In Section 2 we discuss the numerical setup of our simulation, including the structure of the donor star and the heat added to the donor's envelope to initiate MT. Section 3 describes our analysis methods for characterizing the MT rates, the morphology of the outflow, the angular momentum carried by the gas, and the velocity/energies of outflowing gas. Section 4 presents our results for the effects of changing binary mass ratio and envelope heating rate on the outflow's angular momentum and velocity. We compare our results to prior simulations and discuss the implications of our results in Section 5, with a focus on the orbital evolution of the binary and the production of CSM. Finally, we conclude in Section 6.

## 2. Simulation Setup

### 2.1. Grid Setup and Boundary Conditions

We perform Newtonian hydrodynamic simulations using the PLUTO code (A. Mignone et al. 2007, 2012), considering MT between a donor star with core mass $M_1$ and an accretor with mass $M_2$, separated by a distance $a$ in a circular orbit. Our dimensionless simulations have units such that

$$G = M_{\rm tot} = M_1 + M_2 = a = 1,$$

i.e., the orbital separation is 1 and the orbital period is $2\pi$ in these units. The mass ratio is defined as

$$q = \frac{M_2}{M_1} = \frac{\rm accretormass}{\rm donormass}.$$

See Section 2.5 for discussion of scaling our system to physical units for astrophysical binaries.

We adopt third-order Runge–Kutta time-stepping (S. Gottlieb & C. W. Shu 1998) and parabolic spatial integration (A. Mignone 2014). The Riemann solver is set to be `hllc` (E. F. Toro et al. 1994). More details can be found in the PLUTO documentation.

The simulations are performed in 3D spherical coordinates $(r, \theta, \phi)$ in a coordinate frame that is centered on the donor star and corotating with the binary's orbit. The radial grid extends logarithmically between the inner and outer boundaries

$$r_{\rm in} \simeq 0.2, \qquad r_{\rm out} = 5,$$

with the number of grid points chosen such that $\delta r/r \approx 0.01$, leading to about 300 radial grid points. The inner boundary varies slightly (from $r =$ 0.18 to 0.24) with changing $q$, such that the total envelope mass is maintained at approximately 0.05 in code units. The inner radial boundary condition is reflective, and the outer radial boundary condition is outflow, with the additional constraint that the radial velocity $v_r \geqslant 0$ at the outer boundary. The $\phi$ grid extends from 0 to $2\pi$ with 600 uniformly spaced grid points, with resolution chosen such that $\delta\phi \approx \delta r/r$.

The $\theta$ grid extends from 0 to $\pi/2$ with a reflective boundary at $\pi/2$, such that we only simulate the top hemisphere of the binary, as the system is mirror symmetric with respect to the equatorial plane. At $\theta = 0$, we adopt PLUTO's "polaraxis" condition, which accounts for the singularity at the pole. We choose 1 point from $\theta = 0$ to 0.2, 10 points from $\theta = 0.2$ to 0.5, and 100 points from $\theta = 0.5$ to $\pi/2$, choosing lower resolution near the pole to avoid short time steps, but making the resolution in the majority of the grid such that $\delta\theta \approx \delta r/r$.

We define Cartesian coordinates $x$, $y$, and $z$ such that $\hat{x}$ points from the donor star to the companion/accretor, $\hat{x}$ and $\hat{y}$ are in the plane of the orbit, and $\hat{z}$ therefore points along the orbital axis of the binary ($\theta = 0$).

### 2.2. Equations Solved

The equations that are solved in PLUTO are the continuity equations of mass, momentum, and energy as follows:

$$\frac{\partial}{\partial t}\rho + \nabla \cdot (\rho \boldsymbol{v}) = 0 \tag{1a}$$

$$\frac{\partial}{\partial t}(\rho \boldsymbol{v}) + \nabla \cdot (\rho \boldsymbol{v}\boldsymbol{v} + P\boldsymbol{I}) = -\rho\nabla\Phi + 2\rho(\boldsymbol{v} \times \boldsymbol{\Omega}) \tag{1b}$$

$$\frac{\partial}{\partial t}(E_t + \rho\Phi) + \nabla \cdot [(E_t + P + \rho\Phi)\boldsymbol{v}] = 0 \tag{1c}$$

where $\rho$ is the mass density, $P$ is the pressure, $\boldsymbol{v}$ is the velocity in the corotating frame, $E_t = \frac{1}{2}\rho|\boldsymbol{v}|^2 + \rho u_{\rm int}$ is the kinetic +internal energy density, $u_{\rm int}$ is the specific internal energy, and $\boldsymbol{I}$ is the identity matrix. $\Phi$ is the time-independent gravitational potential, which is discussed further below. $\boldsymbol{\Omega} = \Omega\hat{z}$ is the orbital angular frequency (and $\Omega = 1$ in our code units), and $2(\boldsymbol{v} \times \boldsymbol{\Omega})$ represents the acceleration due to the Coriolis force (which does no work and, hence, does not enter the energy equation). $P$ and $u_{\rm int}$ are related by the equation of state (EOS), and we adopt PLUTO's "ideal" EOS

$$P = (\gamma - 1)\rho u_{\rm int} \tag{2}$$

where $\gamma$ is the adiabatic index. This reduces to a polytropic EOS (Equation (4)) if no dissipative terms are included. We use $\gamma = 1.4$ for our fiducial simulations. In our test simulations, we find that changing $\gamma$ does not greatly affect our results, and so we believe that a more realistic EOS would not strongly affect our overall conclusions.

PLUTO's `ENTROPY_SWITCH = SELECTIVE` condition is used, so that the entropy continuity equation (describing the conservation of entropy without dissipative terms) is used to update zones unless in the vicinity of a shock wave, which we found to help eliminate numerical issues in the accretion disk. We also adopt PLUTO's `SHOCK_FLATTENING = MULTID` condition, which reverts to the more diffusive `HLL` Riemann solver upon strong shock detection. This provides additional dissipation near strong shocks and greatly reduces numerical artifacts in our simulations.

### 2.3. Initial Conditions

The majority of the gas is initialized in an envelope atop the core (i.e., above the inner boundary), while a small amount of gas constitutes a low-density ambient background outside the envelope. The core, with mass $M_1$, is treated as a point mass, and the envelope's self gravity is neglected. With the value of the inner radial grid (above), the mass in the envelope is much less than the mass in the core, justifying this assumption. The envelope is assumed to follow a polytropic EOS and is initialized in hydrostatic equilibrium, such that the initial $\rho$ and $P$ are given by

$$\rho = \left(\frac{\gamma - 1}{K\gamma}(\Phi_0 - \Phi)\right)^{\frac{1}{\gamma - 1}} \tag{3}$$

and

$$P = K\rho^{\gamma} \tag{4}$$

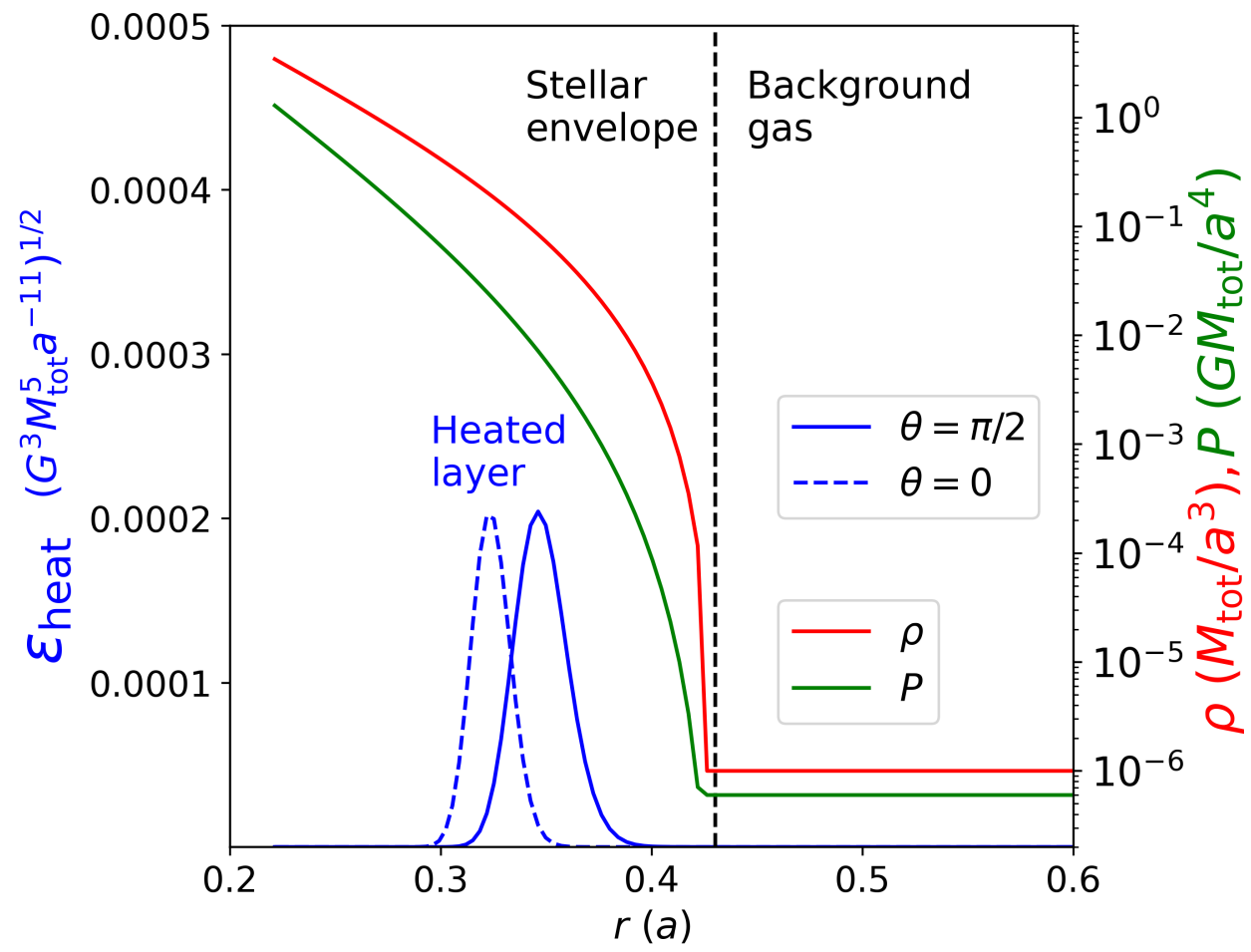

**Figure 1.** Initial conditions and heating for the $q = 0.5$, intermediate heat simulation, q_0.5_mid_heat. Left axis: the steady-state heating profile, power per unit volume, for a line along the pole ($\theta \approx 0$, $\phi = 0$) and a line toward the companion star ($\theta = \pi/2$, $\phi = 0$). Right axis: the initial pressure and density, in code units, along a line toward the companion star ($\theta = \pi/2$, $\phi = 0$), with the vertical dashed line indicating the surface of the star.

where $K$ is the entropy constant, $\Phi$ is the external potential acting on the gas, and $\Phi_0$ is the potential evaluated at the surface of the star.

The potential $\Phi$ is fixed to the Roche potential,

$$\Phi(\boldsymbol{r}) = -\frac{M_1}{|\boldsymbol{r}|} - \frac{M_2}{(|\boldsymbol{r} - a\hat{x}|^2 + \epsilon^2)^{1/2}} - \frac{(x - \mu a)^2 + y^2}{2} \tag{5}$$

where $\boldsymbol{r}$ is the coordinate vector, $\mu = M_2/M_{\rm tot}$, $\epsilon$ is a Plummer softening length, and our choice of units entails the angular orbital frequency is $\Omega = 1$ and the orbital separation is $a = 1$. We adopt $\epsilon = 0.05$ for our fiducial simulations, but vary the value in Appendix C.2. The secondary $M_2$ is a point mass that could represent a stellar companion or a compact object. Its mass remains constant over the course of the simulation.

Throughout the course of the simulation, the center of mass (COM) is fixed and is determined by $M_1$ and $M_2$ but not the mass in the gas. The binary orbit separation $a$ is also held constant in order to focus on steady-state stable MT. In our analysis, we use the Roche lobe calculator of D. A. Leahy & J. C. Leahy (2015) to determine the location of the Lagrange points for different mass ratios.

Note that when implementing the potential $\Phi$, we are using a corotating frame, and PLUTO therefore internally includes the centrifugal force. However, because our origin is not at the COM, we must "manually" supply the $\mu a x$ portion of the centrifugal term (right term of Equation (5)). Therefore, $\Phi_{\rm setup}$, given in the `BodyForcePotential` section of PLUTO, is actually

$$\Phi_{\rm setup} = -\frac{M_1}{|\boldsymbol{r}|} - \frac{M_2}{(|\boldsymbol{r} - a\hat{x}|^2 + \epsilon^2)^{1/2}} + \mu a x. \tag{6}$$

The star is initialized underfilling its Roche lobe (RL), and the ambient material outside the envelope is initialized with a spatially uniform pressure and density, $P = 6 \times 10^{-7}$ and $\rho = 10^{-6}$. We also set a density floor during the simulations, enforcing a minimum density of $\rho = 10^{-9}$. The right-hand axis of Figure 1 plots the initial density and pressure along the line connecting the stars at $\theta = \pi/2$, $\phi = 0$. The density and pressure increase deeper into the envelope of the donor star, and drop dramatically near the surface, plateauing to the initial values of the background material. The mass in the background material is much less than the mass in the envelope.

From Figure 1, the polytropic density profile drops sharply near the surface and cannot be easily resolved. However, we do not believe resolving the initial surface scale height to be necessary for our simulations, as long as the width of the steady-state L1 MT stream is resolved. Simulations that seek to understand the onset and stability of MT may need a more realistic stellar structure and higher resolution near the surface, but that is beyond the scope of this work.

Initially, we assume perfect corotation with the orbit and all gas velocities are set to zero. We do not include cooling or viscosity in our simulations. Therefore, any accretion luminosity from $M_2$ is neglected. Our simulations may therefore be most applicable when the accretor is a star rather than a compact object. We also do not include radiative transfer or radiative forces on the gas, which may significantly accelerate the gas (Section 3.4) in some cases.

### 2.4. Injected Heat

If no heat is added to the envelope, a star that initially underfills its RL remains in hydrostatic equilibrium and continues to underfill its RL, and no MT occurs. To instigate MT, we inject heat within the interior of the envelope, causing it to slowly expand. We set the power injected per unit volume, $\epsilon_{\rm heat}$, to be a Gaussian function of the Roche potential that is ramped up in time to a steady-state value, such that

$$\epsilon_{\rm heat} \propto \exp\left(\frac{-(\Phi - \Phi_h)^2}{2\Delta\Phi^2}\right) \times \frac{1}{2}\left(\tanh\left(\frac{t - t_0}{\Delta t}\right) + 1\right). \quad (7)$$

Here, $\Phi_h$ is the potential at the center of the heated shell, and $\Delta\Phi$ is the width of the shell.

We assume the width of the heating to be the density scale height

$$\Delta\Phi = \left|\frac{\rho}{d\rho/d\Phi}\right| \quad \text{at} \quad \Phi = \Phi_h, \quad (8)$$

which, given our initial density, implies

$$\Delta\Phi = (\Phi_0 - \Phi_h)(\gamma - 1). \quad (9)$$

The heating rate reaches half its maximum steady-state value at $t_0$, and $\Delta t$ controls how fast the heat is ramped up. We fix $t_0 = 15$ and $\Delta t = 5$ in all of our simulations; they are longer than the dynamical timescale of the initial star but shorter than the Kelvin–Helmholtz (KH) timescale of the heated envelope (see Equation (12) later), so our results do not depend on $t_0$ and $\Delta t$. No heat is injected at equipotentials outside the stellar surface.

The normalization of the heating rate per unit volume in Equation (7) is set by the total heat power added to the envelope

$$L_{\rm heat} = \int dV \epsilon_{\rm heat}. \quad (10)$$

The left axis of Figure 1 shows the spatial heating profile for two latitudes $\theta$, once the heating has been ramped up to its maximum value. The heating profile at $\theta = 0$ peaks at smaller radius than the profile at $\theta = \pi/2$, because the star is less extended at the pole than at the equator and heat is added as a function of equipotential.

**Table 1**
Grid of 3D Simulations, Labeled by Mass Ratio and the Amount of Heat Added to the Envelope

| Simulation | $q$ | $x_0$ | $K$ | $t_{\rm kh}$ | $L_{\rm heat}$ |
|---|---|---|---|---|---|
| q_0.5_low_heat | 0.5 | 0.43 | 0.228 | 960 | 2.0e-6 |
| q_0.5_mid_heat | ″ | ″ | ″ | 525 | 3.6e-6 |
| q_0.5_high_heat | ″ | ″ | ″ | 200 | 9.5e-6 |
| q_025 | 0.25 | 0.48 | 0.261 | 525 | 4.0e-6 |
| q_1 | 1 | 0.38 | 0.187 | 525 | 3.2e-6 |
| q_2 | 2 | 0.32 | 0.142 | 525 | 2.5e-6 |
| q_0.5_low_res | 0.5 | 0.43 | 0.228 | 523 | 3.6e-6 |

**Note.** The free parameters are: accretor/donor mass ratio $q$, maximum stellar surface radius $x_0$ (along the $+x$-axis), entropy constant $K$, and heating luminosity $L_{\rm heat}$. $t_{\rm kh}$ is the KH timescale for the heated layer. For all simulations, we adopt polytropic index $\gamma = 1.4$ and mass exterior to the center of the heated layer $M_{\rm ext} \approx 1.0 \times 10^{-3}$. The q_0.5_low_res simulation is nearly identical to q_0.5_mid_heat, but has a lower spatial resolution (see the text).

For comparison between simulations, it is also useful to calculate the exterior mass to the peak of the heating profile, $M_{\rm ext}$, and the KH timescale, $t_{\rm kh}$, for the envelope to expand. We define these as

$$M_{\rm ext} = M(\Phi > \Phi_h), \quad (11)$$

and

$$t_{\rm kh} = \frac{M_1 M_{\rm ext}}{R_h L_{\rm heat}}, \quad (12)$$

where $R_h$ is the approximate radius corresponding to $\Phi_h$, which is roughly circular inside the donor star. See Table 1 for the values of $t_{\rm kh}$ chosen in our simulations, in code units. We note from Table 2 that the timescale over which the envelope expands may be shorter than the thermal or nuclear timescale for a real star to expand and initiate MT. However, the main goal is to simulate high MT rates, and the MT rates in Table 2 are only slightly larger than those expected to occur in massive star binaries (e.g., P. Marchant et al. 2021; S. C. Wu & J. Fuller 2022; J. Klencki et al. 2025), which can easily exceed $10^{-3}\,M_\odot\,{\rm yr}^{-1}$.

### 2.5. Varied Parameters

The initial structure of the polytropic envelope is controlled by six parameters: $a$, $M_1$, $M_2$, $K$, $\gamma$, and $\Phi_0$. These parameters are not independent and can be reduced by one degree of freedom—in this case, we calculate $K$ based on the other parameters. Additionally, by working in units where $M_1 + M_2 = 1$ and $a = 1$, we further reduce the number of parameters by two, leaving only three free parameters. We use the Roche_tidal_equilibrium code[4] (W. Lu 2025) to calculate the hydrostatic profile of a star in a tidal potential. Specifically, we calculate the value of $K$ for a given $q = M_2/M_1$, $\gamma$, and $\Phi_0$. Instead of the surface potential $\Phi_0$, we use the surface position $x_0$ of the star along the $+x$-axis toward the companion, which is interchangeable with $\Phi_0$.

[4] https://github.com/wenbinlu/Roche_tidal_equilibrium

**Table 2**
Conversion between Code Units to Physical Units for Four Example Binaries

| | | Binary Example | | | |
|---|---|---|---|---|---|
| Quantity | Code | A | B | C | D |
| $M_1$ | 2/3 | 10 $M_\odot$ | 10 $M_\odot$ | 10 $M_\odot$ | 2.8 $M_\odot$ |
| $M_2$ | 1/3 | 5 $M_\odot$ | 5 $M_\odot$ | 5 $M_\odot$ | 1.4 $M_\odot$ |
| $a$ | 1 | 0.1 au | 0.22 au | 1.04 au | 0.03 au |
| $R_{\rm L1}$ | 0.442 | 9.5 $R_\odot$ | 21.3 $R_\odot$ | 98.8 $R_\odot$ | 3.1 $R_\odot$ |
| $P_{\rm orb}$ | $2\pi$ | 3 days | 10 days | 100 days | 1 day |
| $L_{\rm heat}$ | 3.6e-6 | 3.5e39 erg s$^{-1}$ | 4.7e38 erg s$^{-1}$ | 1.0e37 erg s$^{-1}$ | 2.5e39 erg s$^{-1}$ |
| $t_{\rm kh}$ | 525 | 250 days | 830 days | 8330 days | 85 days |
| $\dot{M}$ | 0.6e-5 | 6.9e-2 $M_\odot$ yr$^{-1}$ | 2.1e-2 $M_\odot$ yr$^{-1}$ | 2.1e-3 $M_\odot$ yr$^{-1}$ | 5.6e-2 $M_\odot$ yr$^{-1}$ |
| $\bar{v}'_{r,\rm inertial}$ | 0.15 | 55 km s$^{-1}$ | 37 km s$^{-1}$ | 17 km s$^{-1}$ | 51 km s$^{-1}$ |

**Note.** The numbers in code units are parameters and approximate results for the $q = 0.5$, intermediate heat simulation, q_0.5_mid_heat. $M_1$ and $M_2$ are the stellar masses, $a$ is the orbital separation, $R_{\rm L1}$ is the volume-averaged Roche radius of the donor, $P_{\rm orb}$ is the orbital period, $L_{\rm heat}$ is the luminosity added to the donor's envelope, and $t_{\rm kh}$ is the expansion timescale of the envelope (Equation (12)). $\dot{M}$ is the approximate MT rate (discussed in Section 3.2), and $\bar{v}'_{r,\rm inertial}$ is the characteristic radial velocity of an outflow (discussed in Section 3.4). The example binaries have their $M_1$ and $a$ chosen as values appropriate for astrophysical binaries, and all other quantities are then calculated from code units using $M_1$, $a$, and $G$, the gravitational constant (see Equation (13) for an example).

In our simulations, we fix the polytropic index $\gamma = 1.4$ and $x_0/R_{L1} \approx 0.75$ (i.e., not changing the fraction the star fills its RL), and hence, the initial conditions in different cases are only controlled by the mass ratio $q$. From test simulations performed in 2D polar coordinates, we found changes in the surface position $x_0$ not to affect our results. We also tested changing $\gamma$ in 2D simulations, with values of 4/3, 1.4, and 5/3, and found the effects to be subdominant compared to the effect of changing $q$. Therefore, we consider four mass ratios of $q =$ 0.25, 0.5, 1, and 2 as our initial conditions for the 3D simulations.

The dynamical evolution of the system is controlled by two more parameters: the heating luminosity $L_{\rm heat}$ and central position of heat deposition $\Phi_h$ (and the thickness of the heated layer $\Delta\Phi$ is a derived parameter given by Equation (9)). For a given $q$, we determine $\Phi_h$ by taking a *fixed* exterior mass to the $\Phi_h$ layer $M_{\rm ext} \approx 10^{-3}$ in code units (Equation (11)), which determines the depth of the heating inside the envelope, because our 2D test simulations showed it to have no impact, as long as the envelope mass exterior to the heating profile is well resolved and the heating is not too close to the inner radial boundary. Therefore, across different simulations, we only vary the amount of heating injected into the star $L_{\rm heat}$, which controls the mass transfer rate.

Table 1 summarizes the varied properties in the grid of simulations that we perform. The top three rows correspond to simulations with fixed $q = 0.5$ and differing magnitudes of heat added ("high," "mid," "low"). The following three rows correspond to different mass ratios $q$, but keeping heating approximately fixed to "mid" heating values, such that $t_{\rm kh}$ is kept at 525. Finally, we include a low spatial resolution version of one of the simulations, the results of which are discussed in Appendix C. This low-resolution simulation has $\delta r/r = 0.015$ instead of 0.01 (Section 2.1), and correspondingly lower resolution in $\theta$- and $\phi$-coordinates based on using approximately cubic cells.

In the plots that follow, we in general use the default case of intermediate heating, $q = 0.5$ simulation, q_0.5_mid_heat, when showing simulation snapshots.

Because we work in dimensionless code units, we scale our simulations to physical units in example binaries in Table 2. As an example, to convert from velocity in code units $v_{\rm code}$ to a physical value $v$, we use

$$v = v_{\rm code} \times \sqrt{\frac{G(M_1 + M_2)}{a}}. \tag{13}$$

We see that the typical MT rates in the simulations are larger than normally encountered in stellar evolution, and they do correspond to the regime of W. Lu et al. (2023) in which the gas cools inefficiently and our assumption of no cooling is reasonable. Although we are agnostic to the mechanism driving the expansion of the donor star, the heating rates are consistent with those expected for wave heating (e.g., S. Wu & J. Fuller 2020), within the considerable uncertainties.

## 3. Analysis

### 3.1. Flow Morphology

We run the simulations until significant mass transfer has occurred and a quasi-steady state has been reached, where calculated quantities such as the MT rate (Section 3.2) are approximately constant. For the simulation q_0.5_mid_heat, Figure 2 plots two density slices at a late simulation time ($t = 600$), when the simulation has reached a quasi-steady state. One slice lies in the equatorial plane (top panel) and the other is a meriodional slice (bottom panel) corresponding to the $x$–$z$ plane, which is orthogonal to the orbital plane and includes the companion star $M_2$. The density is plotted out to the maximum radius, $r_{\rm out} = 5$. The core of the donor corresponds to the central white-colored region below the inner boundary, which is not part of the computational domain.

The equatorial slice demonstrates the accretion disk around the companion and tails of material outflowing approximately from L2 and L3 points, which then merge to create a broad spiral outflow. The meriodonal slice shows how gas in the accretion disk and outflow is concentrated near the equator, with the density falling off toward the pole. The gas morphology is discussed further in Figures 3–5.

Near the start of the simulation, the heat added to the donor's envelope causes it to slowly expand on a thermal time $t_{\rm kh}$ given by Equation (12) (much longer than its dynamical time) such that it remains close to hydrostatic equilibrium. The outer regions of the donor's envelope expand, but the inner layers below the heated region remain almost unchanged in terms of mass, density, and pressure. Once the star's surface reaches its RL equipotential, mass begins to flow through the inner Lagrange point, L1, toward the companion. Figure 3 demonstrates the progression of the MT by plotting the mass density in the equatorial plane. Note that $t = 2\pi$ represents one orbit of the binary, and the star's dynamical time (for this mass ratio) is $t_{\rm dyn} \approx 0.35$.

By $t = 100$, the star's outer layers have puffed up, and a thin stream of gas flows through L1 and feeds a small accretion disk centered at the companion. At $t = 200$, the accretion disk around the companion has grown substantially. A circumbinary outflow can be seen, as some gas escapes the disk via a stream in the vicinity of L2. Another stream is located on the left side of the donor, in the vicinity of L3. By $t = 600$, the amount of gas outflowing in these streams has increased greatly. The streams from L2 and L3 are expected from prior work (F. H. Shu et al. 1979; O. Pejcha et al. 2016b). Qualitatively, Figures 2 and 3 demonstrate that gas in the disk is able to outflow through outer Lagrange points and escape from the binary.

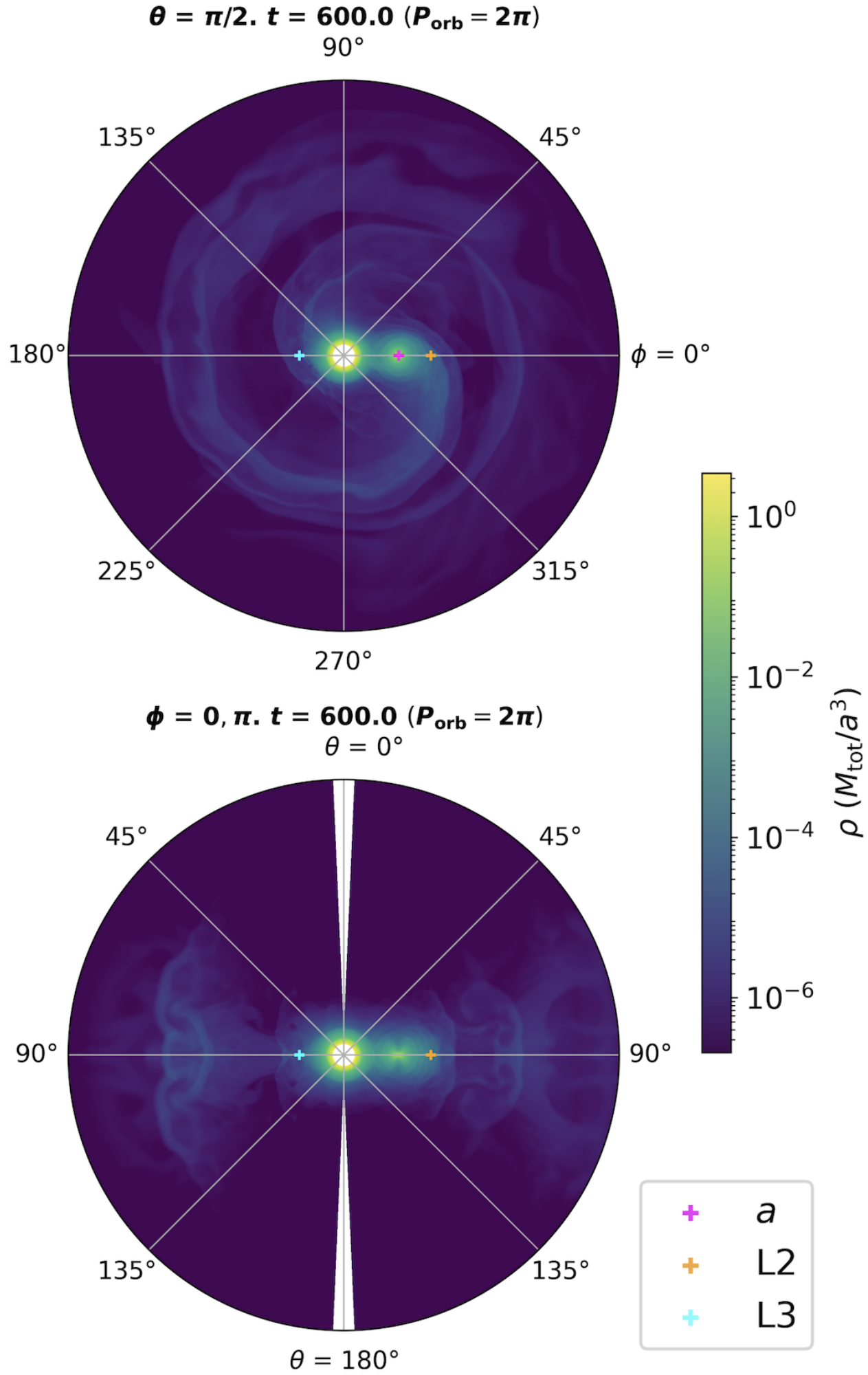

**Figure 2.** Equatorial (top) and meriodional (bottom) slices of the mass density in the intermediate heating, $q = 0.5$ simulation, q_0.5_mid_heat. The domain's full radial extent, out to $r = 5$, is plotted. For the meriodonal slice, we plot the plane that lies between the donor and accretor ($\phi \approx 0$ and $\phi \approx \pi$) and we only simulate $\theta < \pi/2$, so the bottom half of each image is mirrored from the top. The white triangle near $\theta = 0$ is due to low grid resolution at the pole. The legend labels the orbital separation and outer Lagrange points ($a$ is not labeled in the bottom panel for clarity).

Note that Figures 2 and 3 are in a corotating frame, so the outflowing streams lag behind the binary, forming a spiral pattern as gas moves outward and loses angular velocity (F. H. Shu et al. 1979). Spiral internal shocks develop within these streams, as can be seen both in the density and velocity profiles in the circumbinary outflow.

In Figure 4, we overplot velocity vectors in the equatorial plane, at one of the same snapshots as Figure 3. The two panels are the same snapshot, with the upper panel more zoomed in. The fluid flows from the donor star toward the secondary, then circulates the secondary in the accretion disk. Beyond the orbit, the velocity vectors trace the outflowing spiral pattern.

Inside the donor star, the outer layers develop a retrograde flow, opposite to the direction of corotation. There appear to be two effects contributing to this. The first is the expansion of the donor star, as this retrograde pattern begins to develop while the star expands but before significant MT occurs. This can be understood through conservation of angular momentum in the (inertial) lab frame—as the outer layers expand, their rotation frequency drops and hence they lag behind corotation. The second effect is the MT itself, as the flow velocities grow larger as MT occurs. In general, the fluid velocities in the donor star are less than the fluid velocities near L1. Future work should more carefully examine the fluid flows within mass transferring donor stars, as the asynchronous rotation of the layers near the RL may significantly modify the MT transfer rate, as compared to the analytic prediction based on the often adopted corotating hydrostatic configuration.

Figure 5 shows the same simulation and same snapshots, but now for the meriodional plane. The formation of the accretion disk around the companion can be seen edge-on. Although the mass loss tends to be equatorially concentrated, the outflow forms a thick torus due to shock heating and the lack of cooling in our simulations (see also M. MacLeod et al. 2018b).

In the top panel of Figure 6, we plot the density drop-off as a function of polar angle for simulations of varying $q$. The densities are normalized by their value at $\theta = \pi/2$, and averaged over radii, from $r = 2$ to the outer radial boundary at $r_{\rm out} = 5$. The densities are also averaged over $\phi$ and time. In all cases, $\rho$ tends to decrease away from the equator with a roughly exponential profile. With increasing mass ratio (relatively heavier accretor), $\rho$ drops less steeply away from the equator, with $q = 0.25$ showing the most equatorially concentrated outflow, and $q = 2$ showing the least equatorially concentrated outflow. Some curves show slight upticks near $\theta = 0$, but because of the low resolution near the pole, this behavior may not be reliable.

In Appendix A.3, we estimate the average opening angle of outflowing gas $\theta_{\rm outflow,av}$, which corresponds to one density scale height (one $e$-folding from the equatorial value) and quantifies how equatorially concentrated the outflow is. The bottom panel of Figure 6 shows $\theta_{\rm outflow,av}$ plotted versus time for simulations of varying $q$. The steady-state value of $\theta_{\rm outflow,av}$ increases with increasing $q$, from $\approx 10°$ for $q = 0.25$ to $\approx 30°$ for $q = 2$. The physical reason for this trend of $q$ is unclear. This opening angle may change if cooling is included, which would likely reduce the thickness of our disk and potentially change the geometry of the outflow. Therefore, the angles we calculate are likely more accurate at higher MT rates (see Table 2).

Further properties of the gas related to temperature, optical depth, and the photon diffusion time are discussed in Appendices A and A.1. Because we find large optical depths and long photon diffusion times in all of the simulations when they are scaled to high MT rates, our adiabatic assumption and inefficient gas cooling is justified. The possible observational signature of the MT we simulate is discussed in Appendix A.2. Properties of the accretion disk around $M_2$ are discussed in Appendix C.

### *3.2. Mass Transfer Rates*

We calculate the mass-loss rate of the donor by finding the rate of change of mass within a sphere of radius $x_0$, where $x_0$ is the maximum radius of the initial stellar surface, i.e.,

$$\dot{M}_{\rm donor} = \frac{d}{dt}\int_{r=r_{\rm in}}^{r=x_0} \rho\, dV, \tag{14}$$

where $r_{\rm in}$ is the radius of the inner boundary. An alternative way to calculate the MT rate is via a surface flux integral, i.e., $\dot{M}_{\rm donor,alt} = \int \rho \boldsymbol{v} \cdot d\boldsymbol{A}$. This yields similar (but not identical) results. Similarly, the mass-loss rate from the entire computational domain is

$$\dot{M}_{\rm total} = \frac{d}{dt}\int_{r=r_{\rm in}}^{r=r_{\rm out}} \rho dV, \tag{15}$$

where the radial portion of the integral is performed from the inner boundary with radius $r_{\rm in}$ to the outer boundary with radius $r_{\rm out}$.

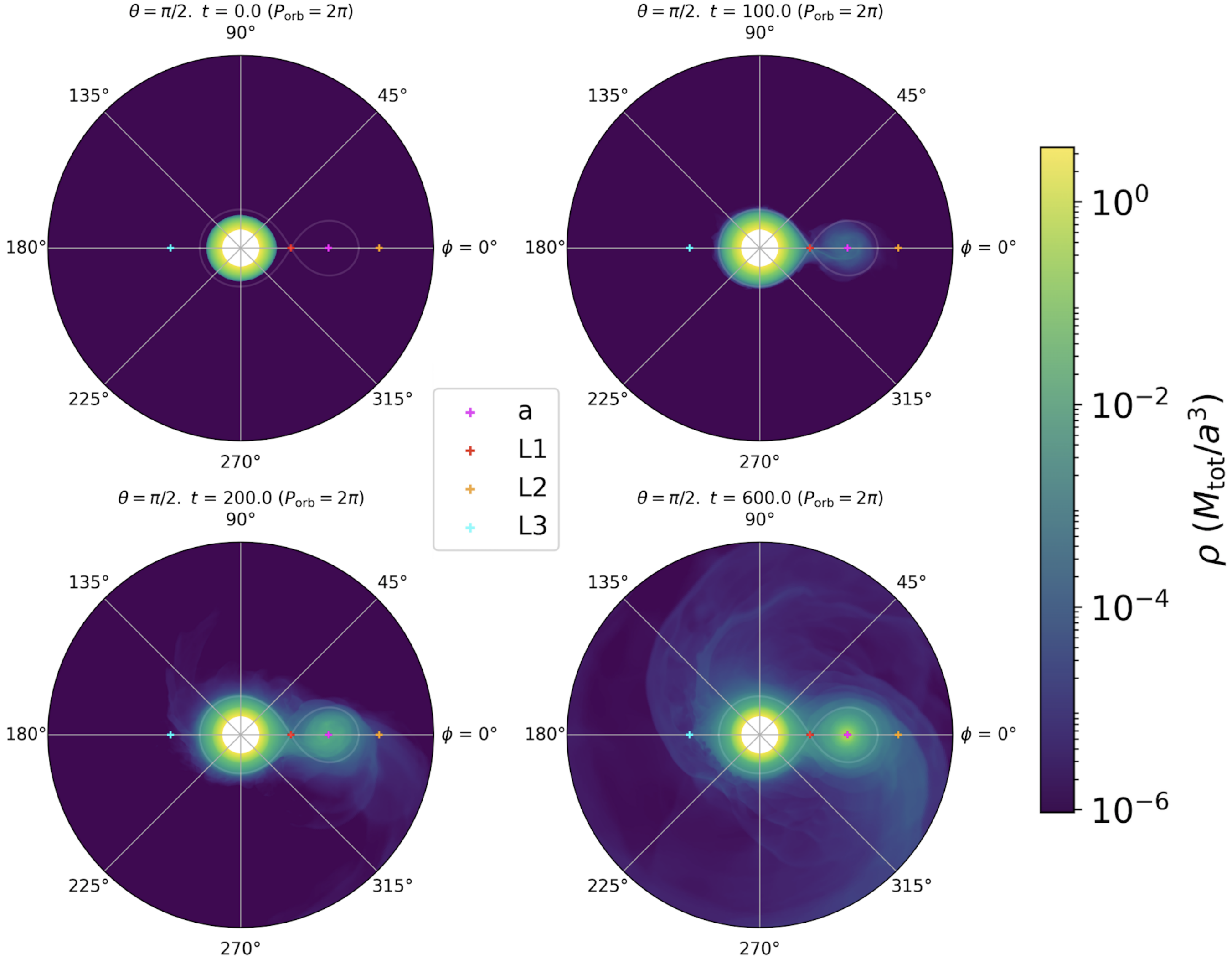

**Figure 3.** Similar to the equatorial slice of Figure 2, but zoomed in, with a maximum $r = 2.2$, and showing the simulation at different time steps. The white contours show the Roche lobes of the binary. The orbital separation and three Lagrange points are labeled in the legend.

Figure 7 plots $\dot{M}_{\rm donor}$ and $\dot{M}_{\rm total}$ over time for the simulation q_0.5_mid_heat. The heating inside the star has been ramped up to its maximum value by $t = 30$, but the MT rate does not stabilize until about $t = 200$, comparable to the thermal time $t_{\rm kh}$ of the donor (Table 1). The rate at which mass leaves the domain also lags behind, stabilizing around $t = 400$, because it takes time for the circumbinary outflow to form and travel to the outer boundary. After $t = 400$, the mass-loss rates are about equal, and the simulation enters a quasi-steady state where the mass-loss rate is nearly constant. Additionally, the mass in the disk is about constant, i.e., the disk is being replenished from the donor star at about the same rate as which mass is lost from the system.

As seen in Figure 3, streams of outflowing gas originate in the vicinity of both L2 and L3. We estimate the mass outflow rate in each of these streams by integrating the mass flux over two surfaces of constant radius but limited in angles. With $r_{\rm L2}/r_{\rm L3}$ as the radial coordinate of L2/L3, one extends around L2 with a radius of $1.1 \times r_{\rm L2}$, from $\phi = -\pi/2$ to $\pi/4$, and the other wraps around L3 from $\phi = \pi/2$ to $5\pi/4$ with a radius of $1.1 \times r_{\rm L3}$. Both surfaces extend from $\theta = 1$ to $\pi/2$. While the exact boundaries of these surfaces are a bit arbitrary, they appear to capture the majority of the mass flow in these two streams. Note that for our $q = 2$ simulation, L2 lies on the side of the donor and L3 on the side of the accretor, i.e., reversed relative to the points plotted in Figure 3.

Figure 8 plots the ratio of the mass flux near L2 to the mass flux near L3, for simulations of varying $q$. The ratio reaches roughly steady-state values by $t = 400$–500, which is similar to when the mass transfer rates reach a quasi-steady state (Figure 7). There is a trend where the ratio substantially decreases with increasing $q$ from $q = 0.25$ to $q = 1$. The ratios for $q = 1$ and $q = 2$ are approximately equal, with a value between unity and 2. Overall, we see that the majority of outflowing gas leaves near the L2 point, even when L2 is on the side of the donor star (and farther away from the accretion disk) in the case of $q = 2$.

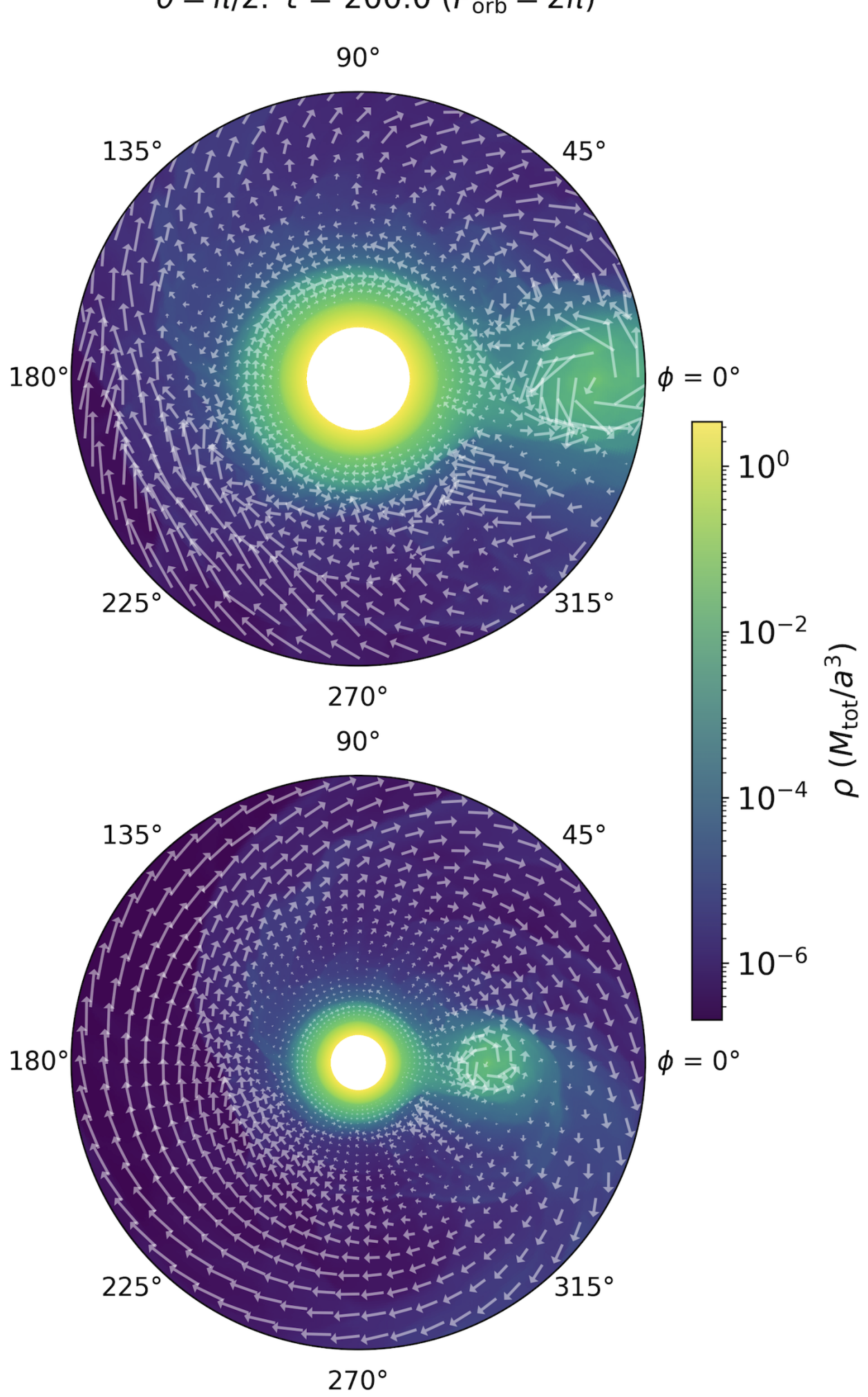

**Figure 4.** Similar to Figure 3, but overplotting the direction of fluid velocities, projected onto the equatorial plane (white arrows). The panels are the same snapshot, but with the bottom panel zoomed out to show larger extent. Velocities are not plotted at every cell, but chosen to illustrate the general behavior.

### 3.3. Angular Momentum Fluxes

We compute the specific angular momentum (AM) carried away by the outflowing gas. The specific AMs (along $\hat{z}$) of the L1, L2, and L3 points are given by $h_{\rm L1} = \Omega(r_{\rm L1} - r_{\rm com})^2$, $h_{\rm L2} = \Omega(r_{\rm L2} - r_{\rm com})^2$, and $h_{\rm L3} = \Omega(r_{\rm L3} - r_{\rm com})^2$, where $r_{\rm L1}$, $r_{\rm L2}$, $r_{\rm L3}$, and $r_{\rm com}$ are the radial coordinates of L1, L2, L3, and the COM, respectively. The specific AMs of $M_1$ and $M_2$ (also along $\hat{z}$) are given by $h_{M_1} = \Omega r_{\rm com}^2$ and $h_{M_2} = \Omega(a - r_{\rm com})^2$. Note again that, for the $q = 2$ simulation, L2 is on the side of the donor star, and L3 on the side of the accretor. The $h$ values for the different $q$ considered in this work are summarized in Table 3.

The specific AM $\boldsymbol{h}$ of an arbitrary fluid element, with coordinate $\boldsymbol{r}$ and velocity $\boldsymbol{v}$ in the simulation frame, is given by

$$\boldsymbol{h} = (\boldsymbol{r} - \boldsymbol{r}_{\rm com}) \times (\boldsymbol{v} - \boldsymbol{v}_{\rm com})_{\rm nr} \quad (16)$$

$$= (\boldsymbol{r} - \boldsymbol{r}_{\rm com}) \times (\boldsymbol{v} + \boldsymbol{\Omega} \times (\boldsymbol{r} - \boldsymbol{r}_{\rm com})) \quad (17)$$

where $\boldsymbol{r}_{\rm com}$ and $\boldsymbol{v}_{\rm com}$ are the coordinate and velocity of the binary's COM (in the corotating frame). The subscript "nr" refers to evaluation in a nonrotating frame. Therefore, the term $\boldsymbol{\Omega} \times (\boldsymbol{r} - \boldsymbol{r}_{\rm com})$ provides the velocity transformation from the corotating frame, where the simulations are performed, to a nonrotating frame. We are interested predominantly in the $\hat{z}$ component of the AM, as the orbital axis of the binary is along $\hat{z}$. The rate of AM outflow $\dot{L}_z$ over a surface with area element $d\boldsymbol{A}$ is given by

$$\dot{L}_z = \int_A h_z \rho \boldsymbol{v} \cdot d\boldsymbol{A}. \quad (18)$$

Similarly, the rate of mass flow through the same surface can be written as

$$\dot{M} = \int_A \rho \boldsymbol{v} \cdot d\boldsymbol{A} \quad (19)$$

and the average specific AM $h_{\rm loss}$ of the outflowing material through this surface is the ratio of these two integrals, i.e.,

$$h_{\rm loss} = \frac{\dot{L}_z}{\dot{M}}. \quad (20)$$

Note that Equation (19) uses an area integral over the flux to evaluate the mass-flow rate, instead of a time derivative of a volume integral (see discussion following Equation (14)). This form is chosen to match the form of Equation (18), and the value $h_{\rm loss}$ is more stable over time when evaluating both integrals as surface integrals.

We find that the value of $h_{\rm loss}$ asymptotes with increasing radius, reaching a nearly constant value similar to that of the L2 point (see Appendix B and Figure 15). The values of $h_{\rm loss}$ for simulations of varying $q$ are discussed further in Section 4.1.

### 3.4. Speed and Energy of Outflowing Material

We also compute the velocity and energy of the outflowing gas. Velocities relative to the COM are

$$\boldsymbol{v}_{\rm inertial} = \boldsymbol{v} + \boldsymbol{\Omega} \times (\boldsymbol{r} - \boldsymbol{r}_{\rm com}).$$

The radial component $v_{r,\rm inertial}$ is calculated as

$$v_{r,\rm inertial} = \boldsymbol{v}_{\rm inertial} \cdot \frac{\boldsymbol{r} - \boldsymbol{r}_{\rm com}}{|\boldsymbol{r} - \boldsymbol{r}_{\rm com}|}. \quad (21)$$

We define the mass-weighted angle-averaged velocity $\bar{v}_{r,\rm inertial}$ as

$$\bar{v}_{r,\rm inertial} = \frac{\int \rho v_{r,\rm inertial} dA}{\int \rho dA} \quad (22)$$

where the integral is performed over a sphere at a constant $|\boldsymbol{r} - \boldsymbol{r}_{\rm com}|$. Doing the integral instead over constant $r$ does not change the following results greatly, which is unsurprising

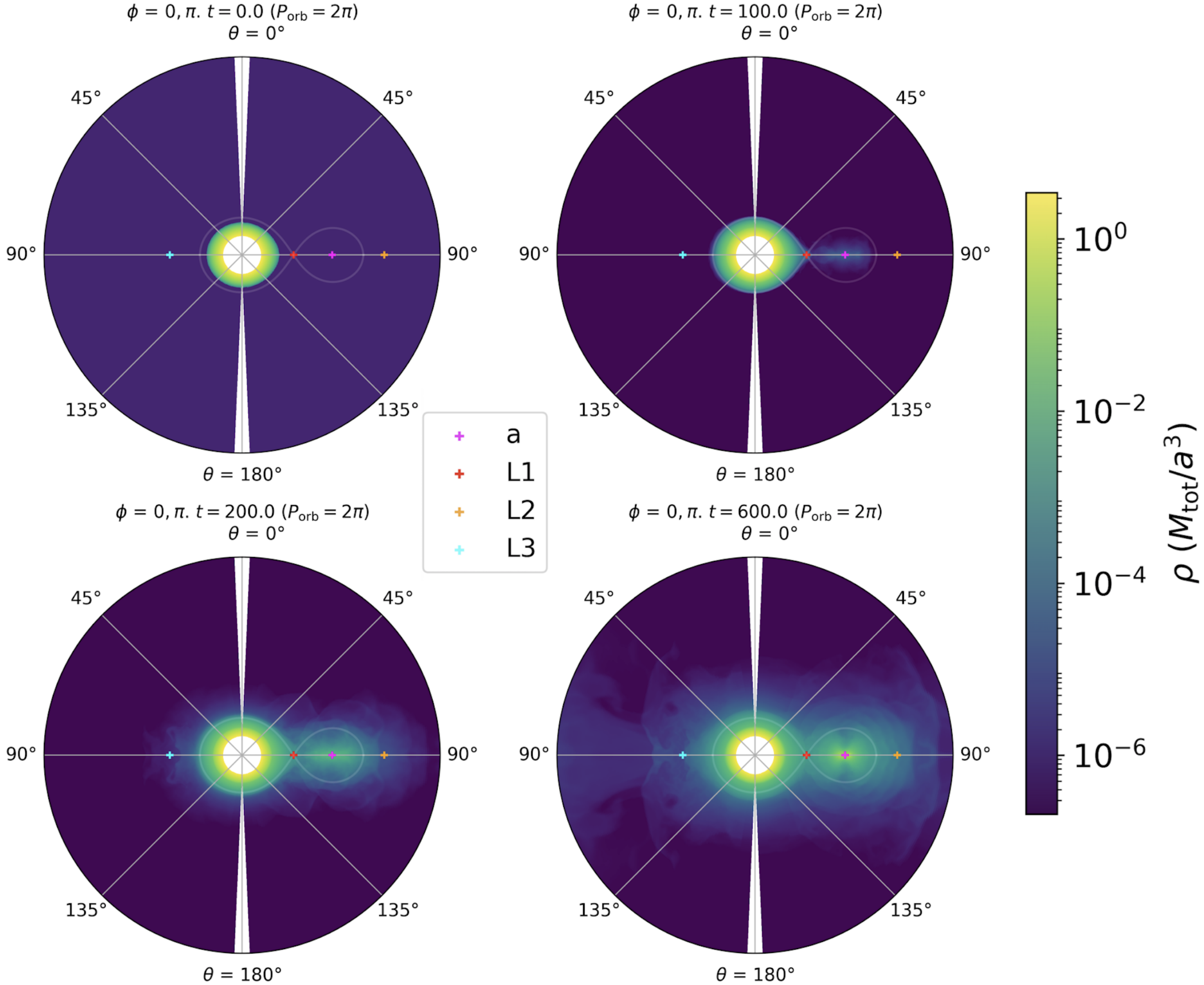

**Figure 5.** Similar to the meriodional slice of Figure 2, but zoomed in, with a maximum $r = 2.2$, and showing the simulation at different time steps. The white contours show the Roche lobes of the binary. The three Lagrange points are labeled in the legend.

given that the COM is located close to the origin, relative to the outer regions of the simulation domain.

The total inertial frame energy per unit mass, $E_{tot}$, is calculated as

$$E_{tot} = \frac{|\boldsymbol{v}_{inertial}|^2}{2} + u_{int} + \Phi_{grav},$$
$$\Phi_{grav} = -\frac{M_1}{|\boldsymbol{r}|} - \frac{M_2}{|\boldsymbol{r} - a\hat{x}|}, \tag{23}$$

where the internal energy is $u_{int} = \frac{P}{(\gamma - 1)\rho}$ and the last term corresponds to the combined gravitational potential of the stars. Similarly, the mass-weighted angle-averaged energy is

$$\overline{E_{tot}} = \frac{\int \rho E_{tot} dA}{\int \rho dA}, \tag{24}$$

where again the integral is performed over constant $|\boldsymbol{r} - \boldsymbol{r}_{com}|$. The Bernoulli parameter $B$ is given by

$$B = E_{tot} + \frac{P}{\rho} \tag{25}$$

and the Jacobi constant $J$ is given as the sum of the Roche potential and the kinetic energy in the corotating frame

$$J = \frac{1}{2}v^2 + \Phi. \tag{26}$$

It is possible to show that the Jacobi constant is also given by $J = |\boldsymbol{v}_{inertial}|^2/2 + \Phi_{grav} - \Omega h_z$, where $h_z$ is the $z$-component of the specific AM. This implies that, as fluid elements gain AM from the binary's torque, they can gain orbital energy as they move out to larger radii. We calculate the corresponding mass-weighted angle-averaged values $\bar{B}$ and $\bar{J}$ similar to $\overline{E_{tot}}$.

Figure 9 shows the density, speed, and energetics of the outflow as a function of distance from the COM, for a typical simulation snapshot. The top panel shows that the angle-averaged density, $\bar{\rho}$, approximately follows a $|\boldsymbol{r} - \boldsymbol{r}_{com}|^{-2}$ dependence, as expected for a constant mass-loss rate and outflow velocity. However, unlike a spherically symmetric wind, the outflow density is concentrated near the equatorial plane. In addition, the kinetic energy of the outflow is dominated by the azimuthal velocity, as expected for a disk-like outflow.

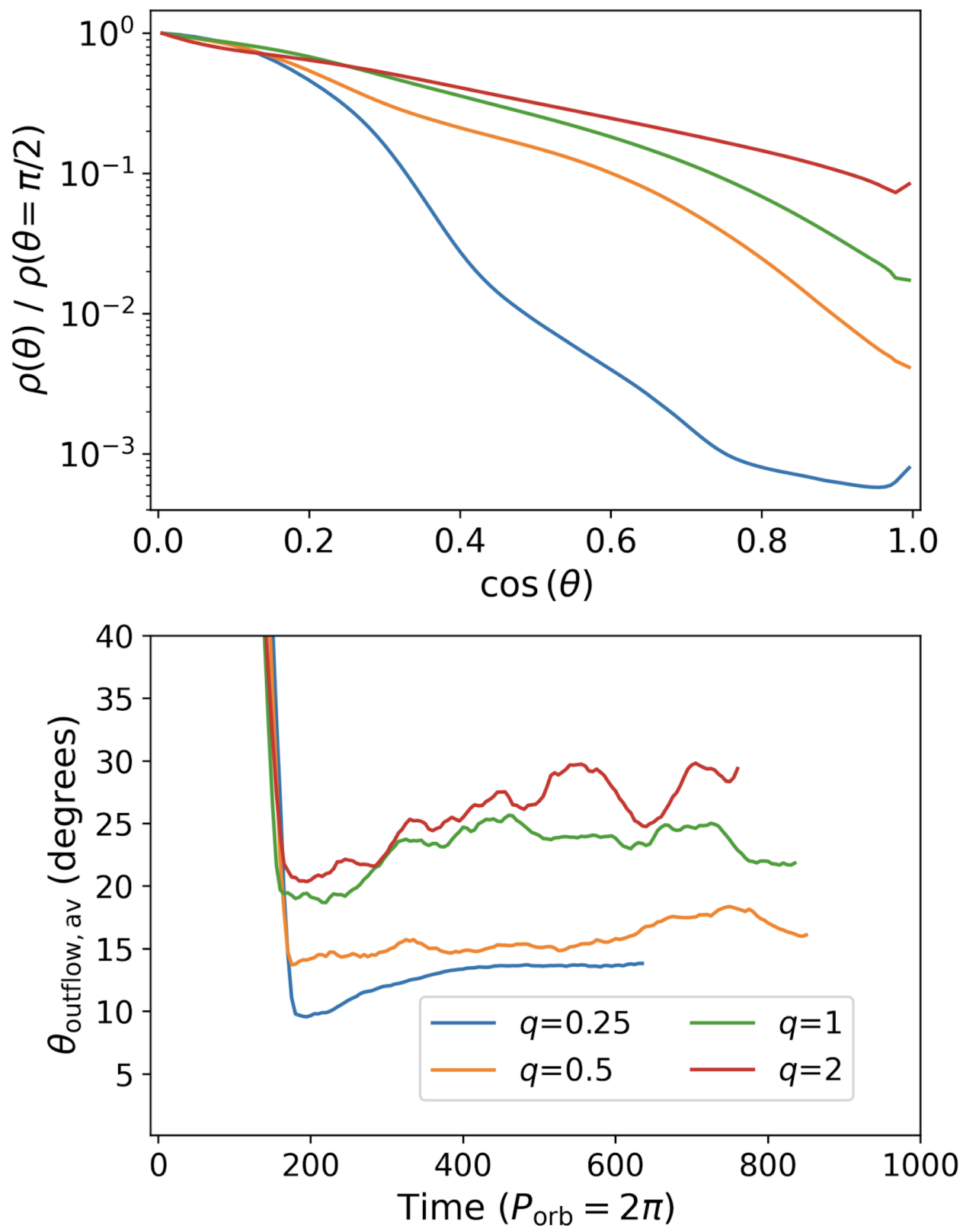

**Figure 6.** Top panel: the average outflow density (normalized by its value at the equator) as a function of angle from the equatorial plane, for simulations of varying $q$ (simulations q_025, q_0.5_mid_heat, q_1, q_2). The $x$-axis corresponds to $\theta$ extending from $\pi/2$ to 0 (equator to pole). Bottom panel: the opening angle of the outflowing gas $\theta_{\rm outflow,av}$, plotted vs. simulation time, for the same simulations. $\theta_{\rm outflow,av}$ corresponds to one density scale height, such that the density has dropped by a factor of $1/e$ compared to its value in the equatorial plane. Each curve is time-smoothed using a rolling average.

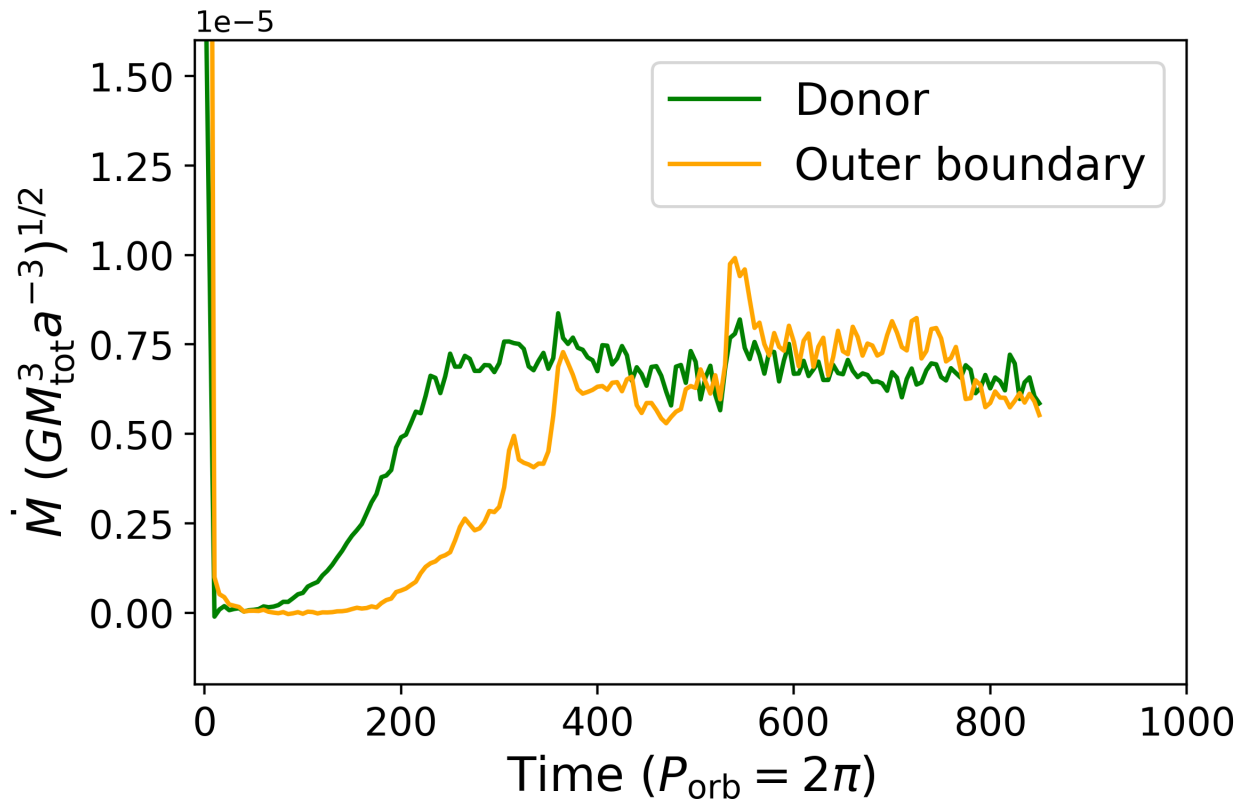

**Figure 7.** The mass-loss rate of the donor star (Equation (14)) and the mass-loss rate of the entire simulation domain (Equation (15)), plotted vs. time for the $q = 0.5$, intermediate heat simulation, q_0.5_mid_heat. A quasi-steady state is reached by $t \simeq 400$.

The bottom panel shows the angle-averaged radial velocity $\bar{v}_{r,\rm inertial}$ and the mass outflow rate $\int dA\ \rho v_{r,\rm inertial}$, where the integral is performed over surfaces of constant $|\boldsymbol{r} - \boldsymbol{r}_{\rm com}|$. We find that $\bar{v}_{r,\rm inertial}$ varies somewhat with radius between

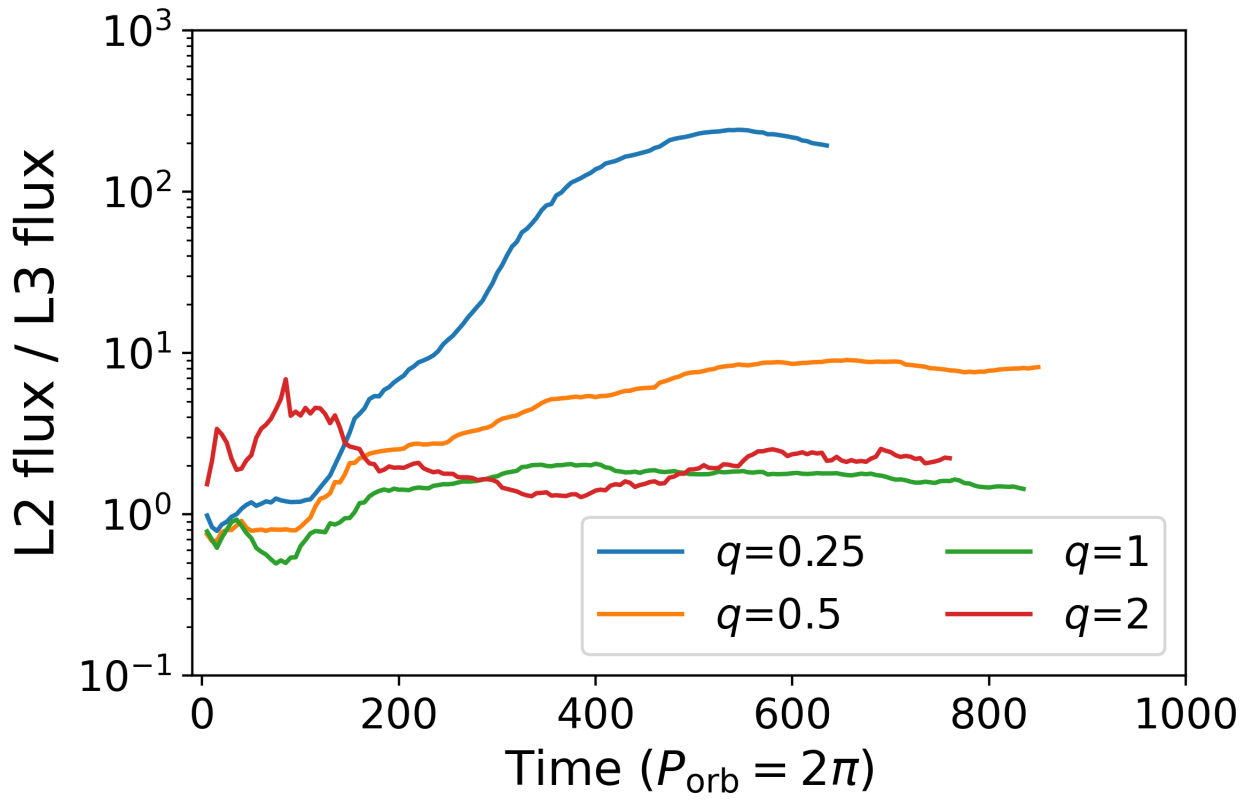

**Figure 8.** The ratio between the mass outflow rates in the stream originating near L2 and in the stream originating near L3. Curves are plotted for varying mass ratio for the same simulations as Figure 6. See the text for a description of the surface areas over which the mass flux is integrated for this calculation. Each curve is time-smoothed using a rolling average.

**Table 3**
For Different Corotating Locations in the Binary, the $z$-component of the Specific AM, $h$, Is Calculated for the Values of $q$ Considered in This Work

| $q$ | $h_{\rm L1}$ | $h_{\rm L2}$ | $h_{\rm L3}$ | $h_{M_1}$ | $h_{M_2}$ |
|---|---|---|---|---|---|
| 0.25 | 0.19 | 1.62 | 1.17 | 0.04 | 0.64 |
| 0.5 | 0.06 | 1.56 | 1.29 | 0.11 | 0.44 |
| 1.0 | 0.0 | 1.44 | 1.44 | 0.25 | 0.25 |
| 2.0 | 0.06 | 1.56 | 1.29 | 0.44 | 0.11 |

**Note.** $h$ is given in units of $\sqrt{GM_{\rm tot}a}$. See the text for the relevant equation to calculate $h$ for each point.

$\bar{v}_{r,\rm inertial}/v_{\rm orb} \sim 0.1$–$0.2$, where $v_{\rm orb} = \sqrt{GM_{\rm tot}/a}$ is the system's orbital velocity. The mass outflow rate is approximately steady with radius.

The top panel of Figure 10 shows the total specific energy and the component energies, each of which is mass-weighted and angle-averaged. Near the outer boundary, the total specific energy $\overline{E_{\rm tot}}$ and Bernoulli parameter $\overline{B}$ slowly increases as the fluid elements climb up the gravitational potential ($\Phi_{\rm grav}$ in Equation (23) becoming less negative). Although the outflow is not exactly ballistic, the Jacobi parameter $\overline{J}$ is approximately constant.

The final fate of the material is unclear. The averaged internal energy $\overline{u_{\rm int}}$ and kinetic energy $\frac{1}{2}\overline{|\boldsymbol{v}_{\rm inertial}|^2}$ appear to asymptote with radius. If this continues at larger radii, the total energy would continue to increase, and the material would become unbound. Physically, this could entail either continued energy input by torques from the binary or work done by pressure gradient forces (the Bernoulli parameter $B$ does not stay constant for a nonsteady flow). However, the tidal torquing far from the binary is minimal—for the simulation snapshot shown in Figure 9, the increase in average specific AM from $|\boldsymbol{r} - \boldsymbol{r}_{\rm com}| = 3$ to $|\boldsymbol{r} - \boldsymbol{r}_{\rm com}| \approx 5$ is $<3\%$ (see also Figure 15).

In the bottom panel of Figure 10, we compare the outflow energetics to that of ballistic trajectories launched near L2. We initialize test particles at a slightly greater $x$-coordinate than L2, and at $y$-coordinates (in units of $a$) of {–0.3, 0, and 0.3} and integrate the equations of motion using a fourth-order Runge–Kutta solver. The trajectories wrap around the binary in a spiral, with trajectories launched at positive $y$ (initially leading the binary's rotation) ultimately becoming more bound compared to those launched at negative $y$ (initially trailing the rotation). This agrees well with the findings of M. MacLeod et al. (2018b) and D. Hubová & O. Pejcha (2019).

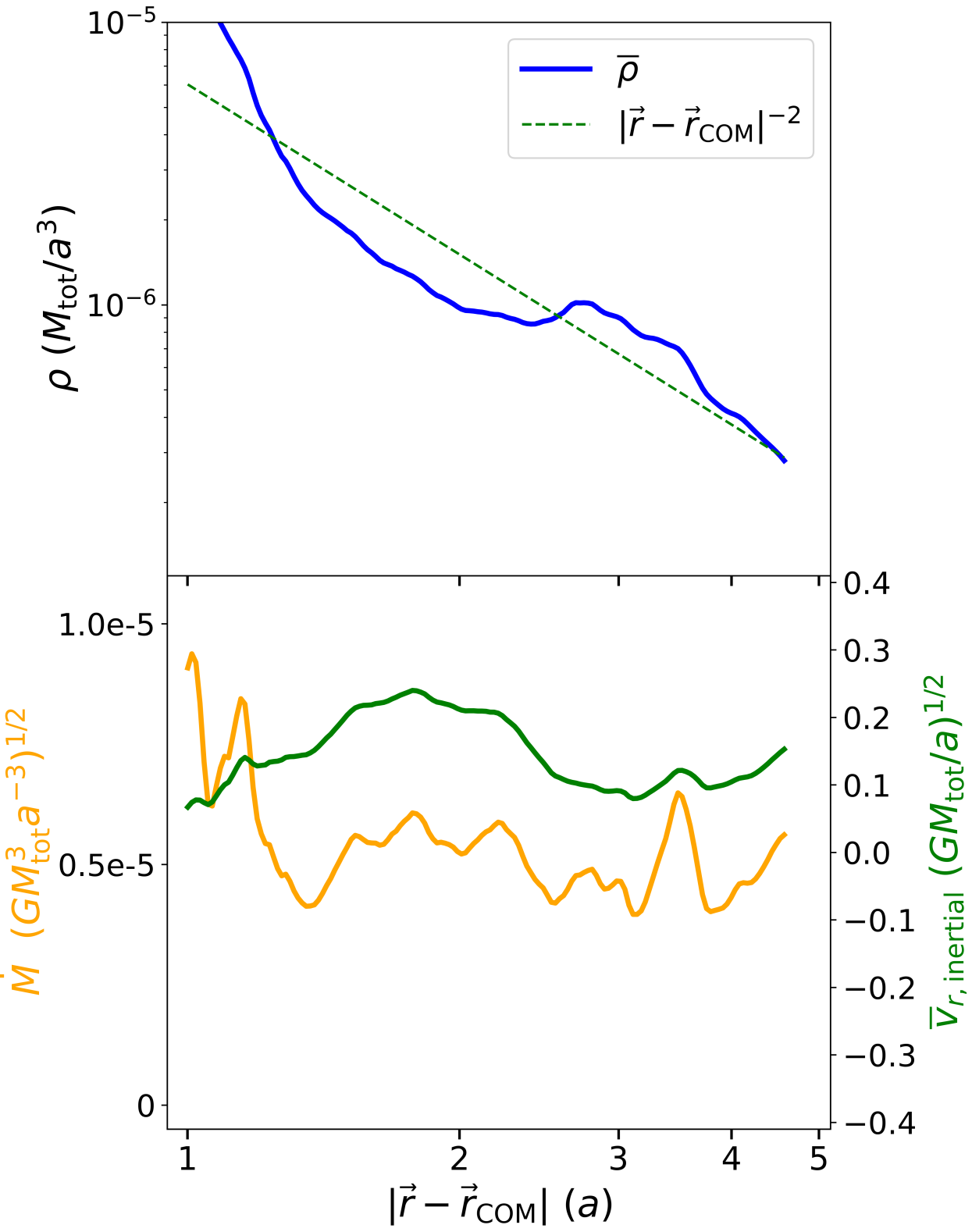

**Figure 9.** Properties of outflowing gas as a function of its distance from the binary COM for the $q = 0.5$, intermediate heat simulation, q_0.5_mid_heat. Top: the angle-averaged density $\bar{\rho}$, in units of $M_{\rm tot}/a^3$, with a $|\boldsymbol{r} - \boldsymbol{r}_{\rm com}|^{-2}$ power-law fit shown to guide the eye. Bottom: the mass-weighted angle-averaged radial velocity $\overline{v}_{r,\rm inertial}$, and the mass outflow rate, evaluated at surfaces of constant $|\boldsymbol{r} - \boldsymbol{r}_{\rm com}|$.

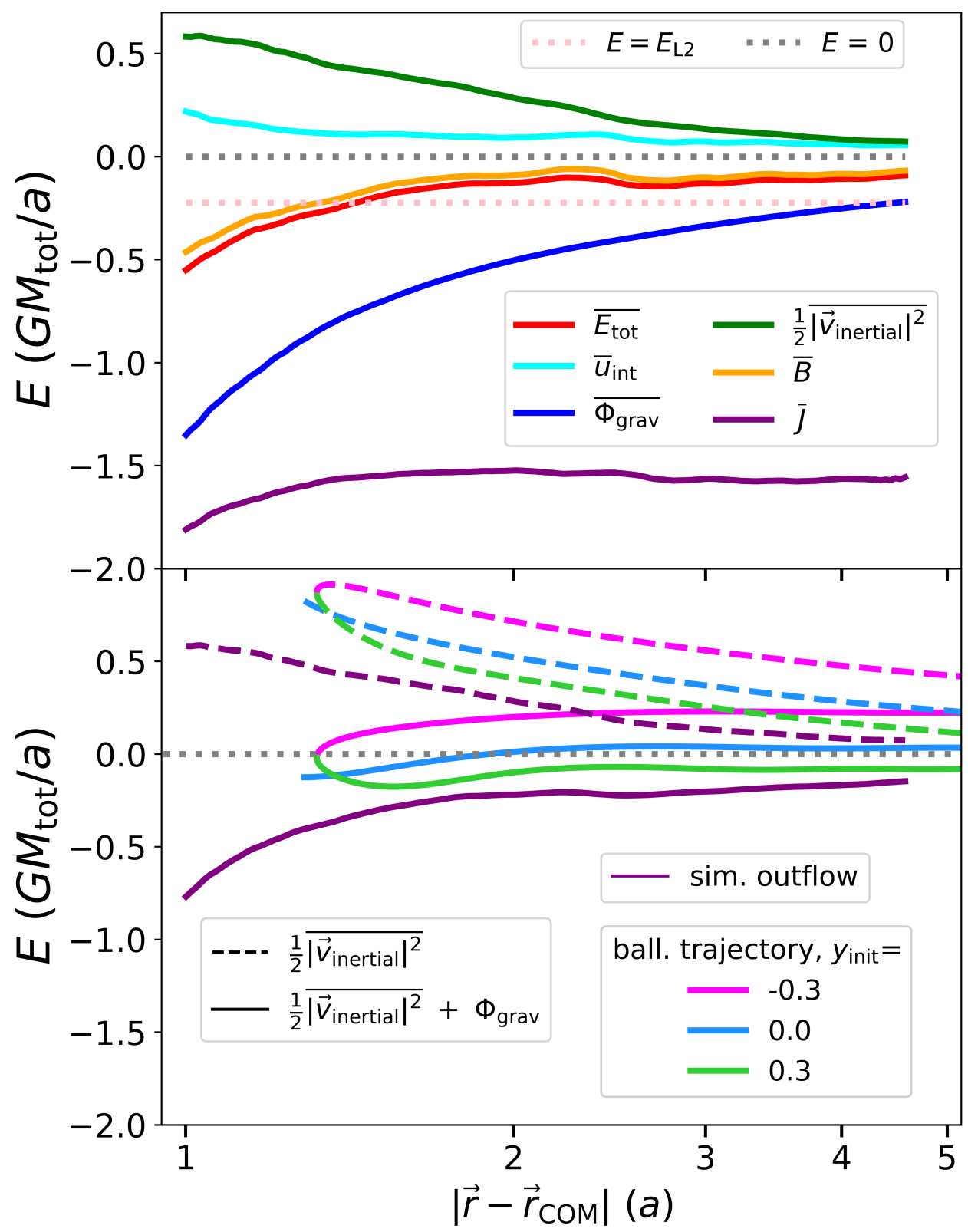

**Figure 10.** Top: the specific total energy $\overline{E_{\rm tot}}$ and each of its component energies (kinetic, internal, and gravitational potential), as seen in Equation (23), as well as the Bernoulli parameter $\bar{B}$ (Equation (25)) and Jacobi constant $\bar{J}$ (Equation (26)). All quantities are mass-weighted and angle-averaged. Dotted lines show where the specific energy equals 0 and $E_{\rm L2}$, the kinetic + gravitational energy of the L2 point. Bottom: a comparison of simulation outflow properties to ballistic trajectories. Purple lines correspond to our simulation outflow. Other colors correspond to ballistic trajectories initialized from rest near L2, but with varying $y$-coordinates from −0.3 to 0.3 (in units of $a$). The kinetic (dashed lines) and kinetic plus gravitational energy (solid lines) are shown for both the outflow and the ballistic trajectories.

The ballistic trajectories have larger kinetic and total energies than that of our outflow. The difference appears to arise from the launching of the outflow. Near the L2 point, the azimuthal velocities in our simulation are only 70%–90% the corotation velocity. The outflow thus begins with lower energy than the ballistic trajectories, and this energy deficit remains approximately constant within our simulation domain. This behavior remains true for simulations of varying $q$.

Note that $u_{\rm int}$ is not included for the energetics of the outflow in the bottom panel of Figure 10. However, the top panel of Figure 10 demonstrates that the addition of $u_{\rm int}$ is not sufficient to unbind the outflow. This does not mean that pressure forces are unimportant. For example, ballistic trajectories initialized near L2 at slightly less than the corotational velocity tend to fall in toward the accretor instead of moving outward. $u_{\rm int}$ is larger than $0.5v_r^2$ near L2, suggesting that pressure forces in particular help launch the outflow near the L2 point, which would not be accounted for in ballistic trajectories. See also Appendix A.4, which discusses how the accretion disk is significantly supported by pressure in addition to rotation.

If the outflow does remain bound, it may form a circumbinary disk that evolves on a viscous time and continues to extract AM from the binary. S. Tuna & B. D. Metzger (2023) modeled the long-term evolution of such disks, finding they could persist for more than $10^4$ yr, and could have a major impact on the appearance of a subsequent supernova explosion (see also A. Ercolino et al. 2025).

Future work, perhaps with an outer boundary farther from the donor, should focus on the behavior of the material far from the binary. However, the effects of radiative cooling due to photon diffusion would become important at large radii (see Appendix A.2 for a brief discussion), so radiative hydrodynamic simulations with realistic opacities are required to fully understand the fate of the circumbinary outflow found in this work.

## 4. Dependence on Simulation Parameters

### 4.1. Effects of Binary Mass Ratio

It is important to understand how the angular momentum and velocity of the outflow depend on the binary parameters such as the mass ratio, $q$. See Section 3.3 for our calculation of AM fluxes and $h_{\rm loss}$. Figure 11 plots $h_{\rm loss}/h_{\rm L2}$ for simulations of different $q$. Here, we only plot $h_{\rm loss}$ averaged over the outermost COM-centered sphere, $|\boldsymbol{r} - \boldsymbol{r}_{\rm com}| = r_{\rm max} \approx 4.65$, because $h_{\rm loss}$ asymptotes with distance from the COM (Figure 15). A steady state is reached at $t \gtrsim 300$, as the curves flatten, while at early times, little MT has occurred. There appears to be a trend that increasing $q$ leads to decreasing $h_{loss}/h_{L2}$, i.e., the material less efficiently extracts AM from the binary. However, the specific AM loss $h_{loss}$ is still substantially larger than the specific AM of the accretor $h_{M_2}$, as shown by the dotted lines of Figure 11 (see also Table 3).

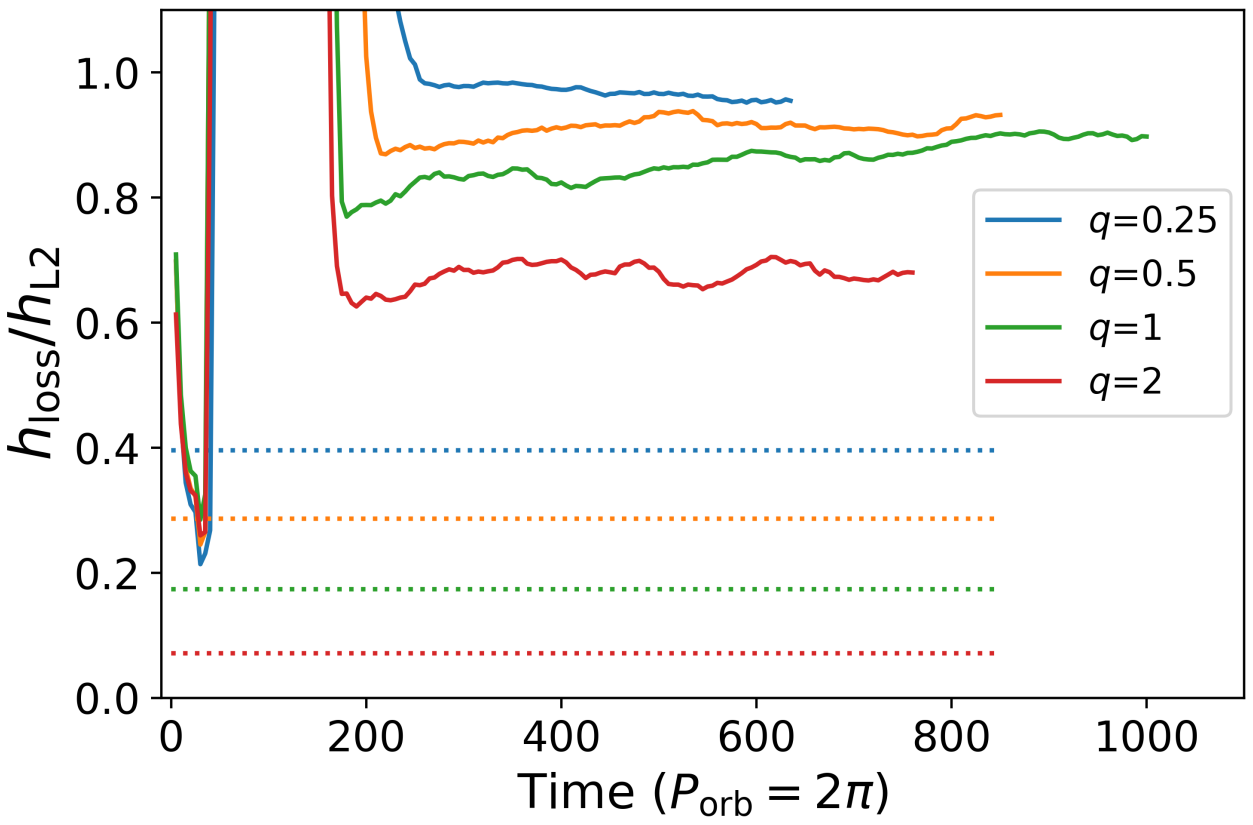

**Figure 11.** The specific AM carried by outflowing gas at $|\boldsymbol{r} - \boldsymbol{r}_{com}| = r_{max}$, for simulations of varying $q$ and intermediate heating (solid lines corresponding to simulations q_025, q_05_mid_heat, q_1, q_2). For each value of $q$, the dotted lines show $h_{M_2}/h_{L2}$, which is constant in time.

The fact that $h_{loss} < h_{L2}$ implies that material has lower AM than if it was corotating at L2 initially (M. MacLeod et al. 2018a). Additionally, although most of the outflowing gas originates in a flow from the disk near L2, some also outflows near L3, which is on the opposite side of the donor star for $q \leqslant 1$ (Figure 8) and on the side of the accretor for $q = 2$. This is one reason why the escaping gas has somewhat smaller specific AM than $h_{L2}$.

Figure 12 compares the radial velocity and energetics of outflows for simulations of differing $q$. See Section 3.4 for our calculation of gas velocities and energies in the inertial frame. The left panel describes how the material near the outer boundary of the simulation domain is bound or unbound, as shown by the mass-averaged Bernoulli parameter $\overline{B}$ over the outermost COM-centered sphere. For all values of $q$, material near the outer radial boundary has negative total energy and Bernoulli parameter, and is thus not yet unbound. However, our simulations do not provide a decisive answer on whether the outflow is unbound (see the discussion in Section 3.4 for uncertainties).

We calculate a mass-averaged radial velocity $\overline{v}'_{r,\mathrm{inertial}}$ in the outer regions of the grid as

$$\overline{v}'_{r,\mathrm{inertial}} = \frac{\int_{r=3}^{r=r_{out}} \rho v_{r,\mathrm{inertial}} dV}{\int_{r=3}^{r=r_{out}} \rho dV}. \tag{27}$$

Note that $\overline{v}'_{r,\mathrm{inertial}}$ involves an integral over angle and radius, whereas $\overline{v}_{r,\mathrm{inertial}}$ of Equation (22) is just integrated over angle. The inner boundary of the integral is chosen to be $r = 3$ to capture the nature of the outflow far away from the two stars. The right panel of Figure 12 plots $\overline{v}'_{r,\mathrm{inertial}}$ for simulations of varying $q$. For $q = 0.5, 1$ and 2, there is a trend where increasing $q$ leads to slightly higher velocities. However, the highest velocities correspond to the $q = 0.25$ simulation. There is a similar trend for $\overline{B}$, where the $q = 0.25$ simulation shows the highest Bernoulli parameter.

The high velocities of the $q = 0.25$ simulation may be problematic, because for a lower accretor mass, the accretor's Roche lobe will be smaller and may be under-resolved in our simulations. Figure 17 shows that a low-resolution $q = 0.5$ simulation also leads to higher characteristic velocities, suggesting our $q = 0.25$ simulation may need higher resolution in order for the outflow velocity to converge. Furthermore, the stream of material passing through L1 may have a comparable thickness as that of the accretion disk, meaning the flow of mass is not as well defined.

However, there are differences in the morphology of the outflow, comparing $q = 0.25$ simulations to those of higher $q$, that are seen in both 3D and 2D polar test simulations. The $q = 0.25$ outflows are heavily dominated by a spiral arm originating at L2, and there is not significant collision with a mass outflow originating at L3. Because these differences are seen in 2D as well, where the MT stream appears to be more resolved, it remains plausible that low $q$ may indeed lead to physical differences in the outflow. We defer further analysis of MT in low $q$ systems to future work, which may require a higher spatial resolution.

### 4.2. Effect of Other Parameters

In 2D test simulations, we did not see a large effect of changing the adiabatic index $\gamma$ on our results. In Appendix C, we investigate the dependence of our results on the envelope heating rate and perform resolution convergence testing.

## 5. Discussion

### 5.1. Comparison to Previous Work

Our work can be compared to several other investigations of binary MT. M. MacLeod et al. (2018a, 2018b) and M. MacLeod & A. Loeb (2020a, 2020b; hereafter, the "M+ papers") investigated pre-common envelope MT and stellar coalescence and, thus, were focused on unstable MT. In our paper, we focus on the steady-state properties of stable MT, and we do not model the coalescence phase.

Despite these differences in focus, there are similarities between our work and the M+ papers. M. MacLeod et al. (2018a) showed mass loss from the binary in spiral flows originating near L2 and L3, and the flow morphology discussed in M. MacLeod et al. (2018b) is similar to ours, including the toroidal structure of the outflow. M. MacLeod et al. (2018b) estimated that the outflowing material is concentrated within 30° of the equator, similar to our results (Figure 6). M. MacLeod et al. (2018b) found that the outflow leaves L2 more tightly bound than corotation would imply, agreeing with our results (Section 3.4).

For a given value of $q$, M. MacLeod et al. (2018b) found that the specific energy and specific AM of the gas asymptotes as a function of radius from the COM. We find similar behavior in regard to specific AM (Figure 15), but our total specific energy is still increasing at our outer radius (Figure 9). This may be a consequence of our simulations having a smaller outer boundary radius. M. MacLeod & A. Loeb (2020a) found that with increasing $q$, the specific AM of outflowing gas is closer to that of L2. They also found a strong dependence of specific AM on the structural index of the donor star $\Gamma_s$ (in our simulations, $\gamma$). In contrast, we find the opposite trend with $q$ (Figure 11) and a weak dependence on $\gamma$. However, M. MacLeod & A. Loeb (2020a) noted that the dependence of specific AM on $\Gamma_s$ is due to the star's response to mass loss and whether the secondary plunges in or remains skimming the star's surface. Because we do not probe this coalescence phase, these differences may be due to modeling stable versus unstable MT. Nonetheless, both works find the specific AM of the gas to be between that of the accretor and L2.

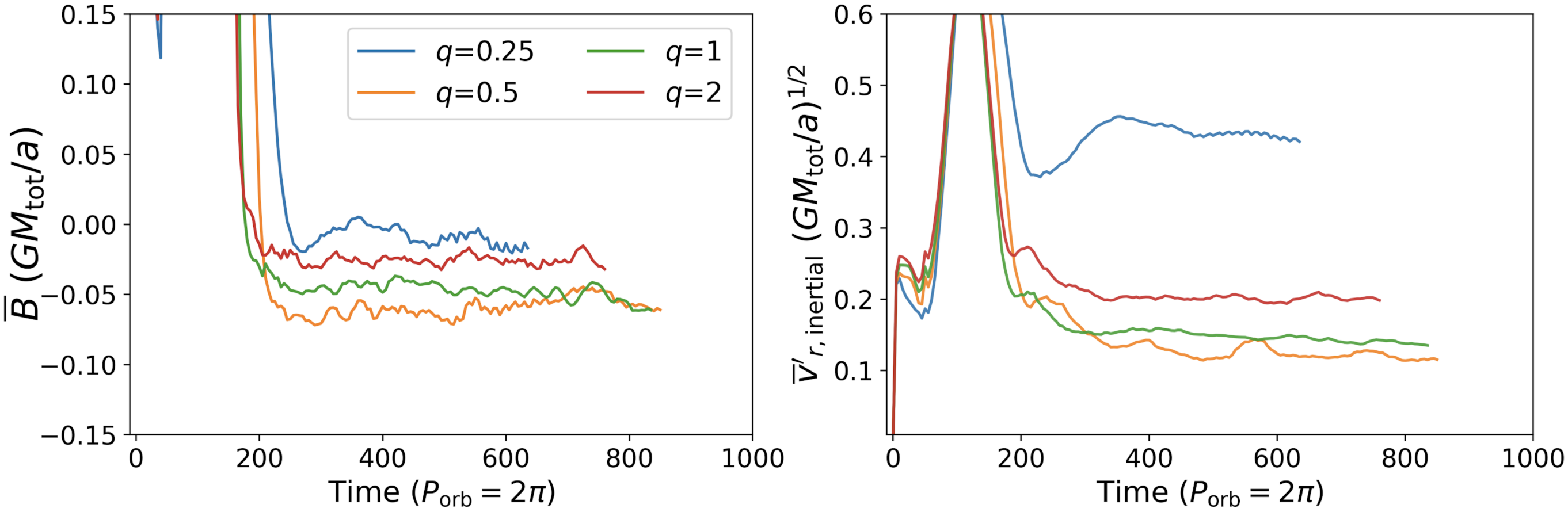

**Figure 12.** Properties related to the energy and velocity of outflowing gas, plotted for the same simulations as in Figure 11. Curves are time-smoothed using a rolling average. Left panel: the Bernoulii parameter $\overline{B}$, in units of $GM_{\rm tot}/a$, averaged over the outermost COM-centered sphere, $|\boldsymbol{r} - \boldsymbol{r}_{\rm com}| = r_{\rm max} \approx 4.65$. Right panel: the volume-averaged radial velocity of the outflow, $\overline{v}'_{r,\rm inertial}$ (Equation (27)), in units of $v_{\rm orb} = \sqrt{GM_{\rm tot}/a}$.

M. MacLeod & A. Loeb (2020b) found that larger $q$ leads to higher radial velocities, which agrees with our results for $q = 0.5$ to $q = 2$, but our highest velocities occur for a low $q$ of 0.25 (Figure 12). However, there are potential caveats to our $q = 0.25$ result (see Section 4.1). M. MacLeod & A. Loeb (2020b) also discussed the effect of changing the gas adiabatic index, $\gamma_{\rm ad}$ (in our simulations, $\gamma$). They found that the gas AM is insensitive to $\gamma_{\rm ad}$, but that the gas radial velocity in the vicinity of the accretor and L2 changes. In our 2D simulations with changing $\gamma$, we find small effects on both gas AM and radial velocity. It is unclear if there is a discrepancy, as we are calculating velocities farther away from the accretor and M. MacLeod & A. Loeb (2020b) referred to velocities near the accretor. If there is a discrepancy, its cause is uncertain, but may be a 2D versus 3D effect, or an effect of the softening potential as discussed in M. MacLeod & A. Loeb (2020b).

Another work that found mass loss from L2 and L3 was Z. Chen et al. (2017), which simulated MT from a pulsating AGB star to a companion. In their simulations where an accretion disk formed, the accretion disk filled its Roche lobe and material flowed out through L2 and sometimes L3, but they did not investigate the angular momentum or the velocity of this outflow.

K. Kadam et al. (2018) performed simulations including the core-envelope structure of both binary components. In the case of their unstable MT simulation, they found mass flow through L2. In their stable MT simulations, the mass accretes directly onto a companion star. The gas never has a chance to form an accretion disk, likely explaining the lack of L2 mass loss in their simulations. Overall, they focused more on the stability of MT and the merger process, rather than the nature of outflowing material.

V. V. Nazarenko et al. (2005) performed simulations of mass transfer in Algol-type binaries, finding mass loss through L2 and L3. By including radiative cooling, they obtained more realistic temperatures than this work (see Appendix A).

J. L. A. Nandez et al. (2014) found L2 mass loss preceding a merger of a giant star and a companion, when dynamical timescale MT occurred. T. A. Reichardt et al. (2019) saw L2 and L3 mass loss in a simulation preceding a common envelope, which formed a bound disk that later interacted with common envelope ejecta. In contrast to these works, we focus on steady-state stable MT, not mergers.

A. Bobrick et al. (2017) performed simulations of mass transfer between a white dwarf and neutron star, finding that at Super-Eddington MT rates angular momentum was efficiently removed through disk winds. They found that mass loss through L2 is not expected for such binaries, but the MT rates they modelled are lower than considered in this work.

Several works have investigated MT in eccentric binaries, where MT occurs episodically near periastron. E. Regös et al. (2005) found that mass escapes through the L2 point for high eccentricities, when the potential of L2 approaches that of L1. Similarly, in the SPH sims of R. P. Church et al. (2009), most of the transferred particles escaped through L2, which has a similar potential to L1 for their binary parameters. C.-P. Lajoie & A. Sills (2010a) noted mass loss through L2 and sometimes L3, depending on the component masses (see also C.-P. Lajoie & A. Sills 2010b).

We note that both R. P. Church et al. (2009) and C.-P. Lajoie & A. Sills (2010a) were able to simulate low MT rates in their eccentric binaries (down to $10^{-10} M_\odot\,{\rm yr}^{-1}$ in the case of R. P. Church et al. 2009), whereas the assumptions of our work demand much higher MT rates. However, neither of these works appear to have included radiative cooling. In contrast, R. A. Booth et al. (2016) performed sophisticated modeling of cooling in their simulations of symbiotic novae while also modeling a low MT rate ($10^{-7} M_\odot\,{\rm yr}^{-1}$). They investigated the interaction of novae with CSM, which differs from our focus on steady-state stable MT. S. Mohamed & P. Podsiadlowski (2012) also simulated low MT rates with cooling in wind RLOF and found that some of the wind leaves through L2 to form a spiral arm. Future work can improve upon ours by including radiative cooling in order to realistically simulate lower MT rates, which are more common than those considered in this work.

Works that predict mass loss through L2, but without modeling MT from a donor star, include F. H. Shu et al. (1979), O. Pejcha et al. (2016a, 2016b), and D. Hubová &

O. Pejcha (2019). While O. Pejcha et al. (2016b) predicted unbound outflows, D. Hubová & O. Pejcha (2019) showed that the energetics of the outflow can depend on initial conditions in the vicinity of L2, such as the positional offset from L2 and initial velocities of the gas. Our work does does not prescribe the initial condition of the outflow near L2, as we instead drive MT from the donor star toward the accretor. Despite this improvement, there remain uncertainties in our work as to whether the material is ultimately bound or unbound (Section 3.4).

Recently, T. Ryu et al. (2025) also simulated steady-state mass transfer, but their simulations focused on flows through the L1 or L2 point and did not include the accretor or circumbinary flows. Their simulations are more useful for understanding the donor response to mass transfer and the width/shape of the L1 stream, while our simulations are more useful for understanding circumbinary outflows and angular momentum loss.

### 5.2. Implications for Binary AM Loss

For nonconservative MT, the AM carried by outflowing gas is extremely important for predicting the orbital evolution (S. S. Huang 1963; R. Willcox et al. 2023). Many prior works, including some binary population synthesis codes, have assumed that gas carries a specific AM $h$ corresponding to that of the accretor. Our finding that $h$ is closer to that of the L2 point means that more angular momentum will be extracted from the binary, likely leading to more dramatic orbital decay during phases of stable mass transfer. We note that our results only apply to high rates of MT ($\dot{M} \gtrsim 10^{-4} M_\odot$ yr$^{-1}$[; see W. Lu et al. 2023), and future work should carefully define where our results apply.

Stable MT has become a leading candidate for the creation of isolated BBHs (e.g., K. Pavlovskii et al. 2017; M. Gallegos-Garcia et al. 2021; J. Klencki et al. 2021; P. Marchant et al. 2021; L. A. C. van Son et al. 2022). Stable MT plus efficient AM loss through L2 alters predicted BBH population properties (e.g., orbital periods, mass ratios, and merger rates) versus stable MT without L2 mass loss (A. Picco et al. 2024), and may either increase or decrease BBH merger rates compared to stable MT without L2 mass loss. This will also change the BBH merger delay time distribution and the BBH spins due to tidal spin-up (e.g., Y. Qin et al. 2018; S. S. Bavera et al. 2020, 2021; L. Ma & J. Fuller 2023; A. Olejak et al. 2024).

We do not evolve the binary's orbital separation due to AM losses, so our simulations at face value cannot be used to determine the stability of mass transfer. Rather, our goal is to understand the quasi-steady state during the evolution of binaries in which mass transfer does remain stable. Future work should self-consistently evolve a binary with AM losses appropriate to mass flow through L2, to better understand its role in stable and unstable MT.

### 5.3. Implications for Circumstellar Material

When mass transfer is occurring at the end of a massive star's life, the circumbinary outflow seen in this work results in CSM that will interact with the supernova ejecta. B. D. Metzger & O. Pejcha (2017) investigated a related scenario involving the interaction of stellar merger ejecta with circumbinary material ejected before the merger. Future work should build upon B. D. Metzger & O. Pejcha (2017) by investigating the interaction of supernova ejecta with circumbinary material, with properties informed by our 3D simulations. Predicting supernova lightcurves and spectra to compare to observations could then determine what fraction of interacting supernova arise from this scenario.

To perform this analysis, we would need to estimate the radial extent and density of the CSM when the supernova occurs. We have already shown that the outflow density approximately follows an $r^{-2}$ dependence (Figure 9). We can estimate the CSM radius $R$ given a time $t_{\rm delay}$ between the onset of MT and the explosion of the star. A simple estimate, based on the asymptotic $v_r \approx 0.2\sqrt{GM_{\rm tot}/a}$ we find (Section 3.4 and Figure 12), is

$$R \approx 0.2 \times t_{\rm delay}\sqrt{GM_{\rm tot}/a} \approx 2 \times 10^{15}\ {\rm cm} \left(\frac{t_{\rm delay}}{10\ {\rm yr}}\right)\left(\frac{M_{\rm tot}}{15 M_\odot}\right)^{0.5}\left(\frac{a}{0.1\ {\rm au}}\right)^{-0.5}. \quad (28)$$

Since the outflow speed is only ~20% that of the orbital speed, the CSM can be significantly more confined than some prior estimates, which assume it is ejected near the orbital speed.

### 5.4. Future Work

As also discussed in M. MacLeod & A. Loeb (2020b), there are numerous physical effects that may complicate the accretion flow around the donor star and ultimately affect the global outflow. These include radiative cooling and magnetic fields, the latter of which governs the disk's accretion onto $M_2$. Additionally, if the accreting secondary is a compact object, it will be a source of energetic feedback on the outer part of the disk. Because we are not including any feedback, our current simulations may be more accurate for stellar companions, as long as they are small enough for an accretion disk to form. Farther from the accretor, dust formation will occur at low temperatures, and radiation pressure on the dust may become important, as modeled in S. Mohamed & P. Podsiadlowski (2007, 2012). Radiation pressure may help the gas to become more unbound from the binary.

Future work should also investigate the role of these circumbinary outflows in supernova interactions (Section 5.3) and decaying binary orbits, especially in regard to creating merging compact binaries through stable MT (Section 5.2). If the accretor is a star, one possibility is that some of its envelope is mixed into the outflow, potentially changing the outflow's composition as compared to that of the donor star. This may lead to interesting CSM compositions that can explain some classes of supernova.

## 6. Conclusion

Episodes of intense and nonconservative mass transfer occur in many binaries containing a massive star. We simulated mass transfer between a donor star and an unresolved secondary to investigate the angular momentum, velocity, and energetics of outflows. Our idealized simulations do not include cooling, and are expected to be most applicable to systems with very large mass transfer rates of $\dot{M} \gtrsim 10^{-3} M_\odot$ yr$^{-1}$. Our 3D simulations are performed using the PLUTO hydrodynamic code, which resolves the flow of mass from the donor, into an accretion disk around the secondary, through the outer

Lagrange point (L2), and into a circumbinary outflow. For our main grid of simulations, we vary the mass ratio $q$, defined as the accretor mass over donor mass.

The main results of our work are as follows:

1. Mass from the accretion disk tends to outflow in a stream originating near L2 and forming a spiral pattern (Figures 2–4). The circumbinary disk outflow has a half opening angle of $\theta = 10°–30°$, increasing with increasing $q$ (Figure 6). The simulation reaches a quasi-steady state where the mass-loss rate of the donor star approximately equals the rate of mass flowing through the outer boundary (Figure 7). Our main results are calculated when this quasi-steady is realized.
2. The outflowing gas has a specific angular momentum $h_{\rm loss}$ that asymptotes with radius to a value slightly less than that of the L2 point, $h_{\rm L2}$ (Figure 15). This means that AM will be efficiently extracted from the binary, which will have important consequences for orbital evolution. The outflow is dominated by mass originating near L2, although a stream near L3 also contributes some mass loss, especially for higher $q$ (Figure 8).
3. $h_{\rm loss}/h_{\rm L2}$ depends moderately on $q$ (Figure 11), with values of approximately {0.95, 0.9, 0.8, 0.65} for $q$ of {0.25, 0.5, 1, 2}. This suggests that lower mass ratios may lead to more efficient AM loss through L2. In all cases, $h_{\rm loss}$ is much larger than the specific AM of the accretor, $h_{M_2}$.
4. For most $q$, the outflow appears to reach an asymptotic radial velocity $\bar{v}_{r,\rm inertial}$ that is between 10% and 20% the orbital velocity (Figures 9 and 12). Velocities may be higher for $q = 0.25$. The outflow's total specific energy remains negative but is slowly increasing, suggesting that some or all of the gas may ultimately be unbound. The density roughly follows an $r^{-2}$ dependence (Figure 9).
5. The adiabatic index of the gas, $\gamma$, and the form of the heating injected into the envelope (Figure 17) do not greatly change the above results for $h_{\rm loss}$ and $\bar{v}_{r,\rm inertial}$ or change the morphology of the outflow.

Although there have been several prior works that have performed hydrodynamic simulations and seen mass outflow through L2 and/or L3, most have been focused on unstable mass transfer leading to stellar mergers. Our measurement of angular momentum carried away during stable but nonconservative mass transfer has major implications for the resulting orbital evolution. These results can be incorporated into population synthesis calculations of stable MT and resulting mergers of binary compact objects (e.g., M. Gallegos-Garcia et al. 2021; P. Marchant et al. 2021). They will also be useful for predicting the properties of CSM following extreme mass loss from supernova progenitors (S. C. Wu & J. Fuller 2022). Many uncertainties remain, and future work should examine cooling processes of the gas, dust formation, MHD processes and energetic feedback from the accretor's disk, flow dynamics within the donor star, and flow dynamics farther from the binary.

## Acknowledgments

We are grateful for support from the NSF through grant AST-2205974. This research benefited from interactions enabled by the Gordon and Betty Moore Foundation through grant GBMF5076. Numerous conversations have helped lead this project to its current state. We thank Sterl Phinney for advice that initiated this project, Morgan MacLeod for help with the simulation setup, and Elias Most for advice on running the simulations on a cluster. We thank Dan Kasen and Tony Piro for useful feedback that generated more points of discussion. We are grateful to the organizers and participants of the 41ST Liége International Astrophysical Colloquium and the 2024 ZTF Theory Network Conference, at which we received advice and questions that led to new analysis. We thank the anonymous referee for providing thoughtful comments, which greatly improved this work. We also thank the data editor for providing feedback.

This work used the Purdue Anvil supercomputer through allocation PHY240274 from the Advanced Cyberinfrastructure Coordination Ecosystem: Services & Support (ACCESS) program, which is supported by U.S. National Science Foundation grant Nos. 2138259, 2138286, 2138307, 2137603, and 2138296.

*Software:* PLUTO, Python, NumPy (C. R. Harris et al. 2020), Matplotlib (J. D. Hunter 2007), SciPy (P. Virtanen et al. 2020), Roche lobe calculator (D. A. Leahy & J. C. Leahy 2015), Roche_tidal_equilibrium[5] (W. Lu 2025).

## Appendix A
## Gas Properties

Figure 13 plots the temperature $T$ of the gas in the equatorial plane for a typical simulation snapshot. $T$ is calculated by assuming the total pressure is a combination of ideal gas and radiation pressure,

$$P = \frac{\rho k_{\rm b} T}{\mu m_\mu} + \frac{1}{3} a_{\rm rad} T^4, \tag{A1}$$

where $k_{\rm b}$ is the Boltzman constant, $m_\mu$ is the atomic mass unit, $a_{\rm rad}$ is the radiation constant, and $\mu$ is the mean molecular weight, assumed to be 0.61 for solar composition. In order to calculate physical values of $T$, $\rho$, and $P$, we adopt the fiducial case of an $M_1 = 10\,M_\odot$, $M_2 = 5\,M_\odot$ binary in a 10 day orbit (Example B in Table 1). However, without even assuming binary parameters, the colors of Figure 13 show the relative values of $T$ at different points in the grid. Because we do not include gas cooling, the temperatures can be quite hot ($T \gtrsim 10^5$ K).

### A.1. Optical Depth Estimates

We calculate the optical depth $\tau$ and photon diffusion time $t_{\rm phot}$ for several radial lines of sight through the gas, where

$$\tau = \int_{r_0}^{r_1} \kappa \rho dr \tag{A2}$$

and

$$t_{\rm phot} = \int_{r_0}^{r_1} \frac{\kappa \rho (r - r_0)}{c} dr \tag{A3}$$

where $r_0$ and $r_1$ are the radial endpoints of the path integral, which is along a constant value of $\theta$ and $\phi$. $\kappa$ is the opacity, which we assume is the Thomson scattering opacity appropriate for solar composition, 0.34 cm$^2$ g$^{-1}$, and $c$ is the speed of light. The value of $\tau$ depends on which path is chosen and the amount of gas it intersects, but we do not include the gas in the donor star's envelope in the integral. Instead, we perform integrals from outside the envelope with $r_0 \approx 0.6$, to the outer boundary, at different values of $\theta$ and $\phi$.

[5] https://github.com/wenbinlu/Roche_tidal_equilibrium

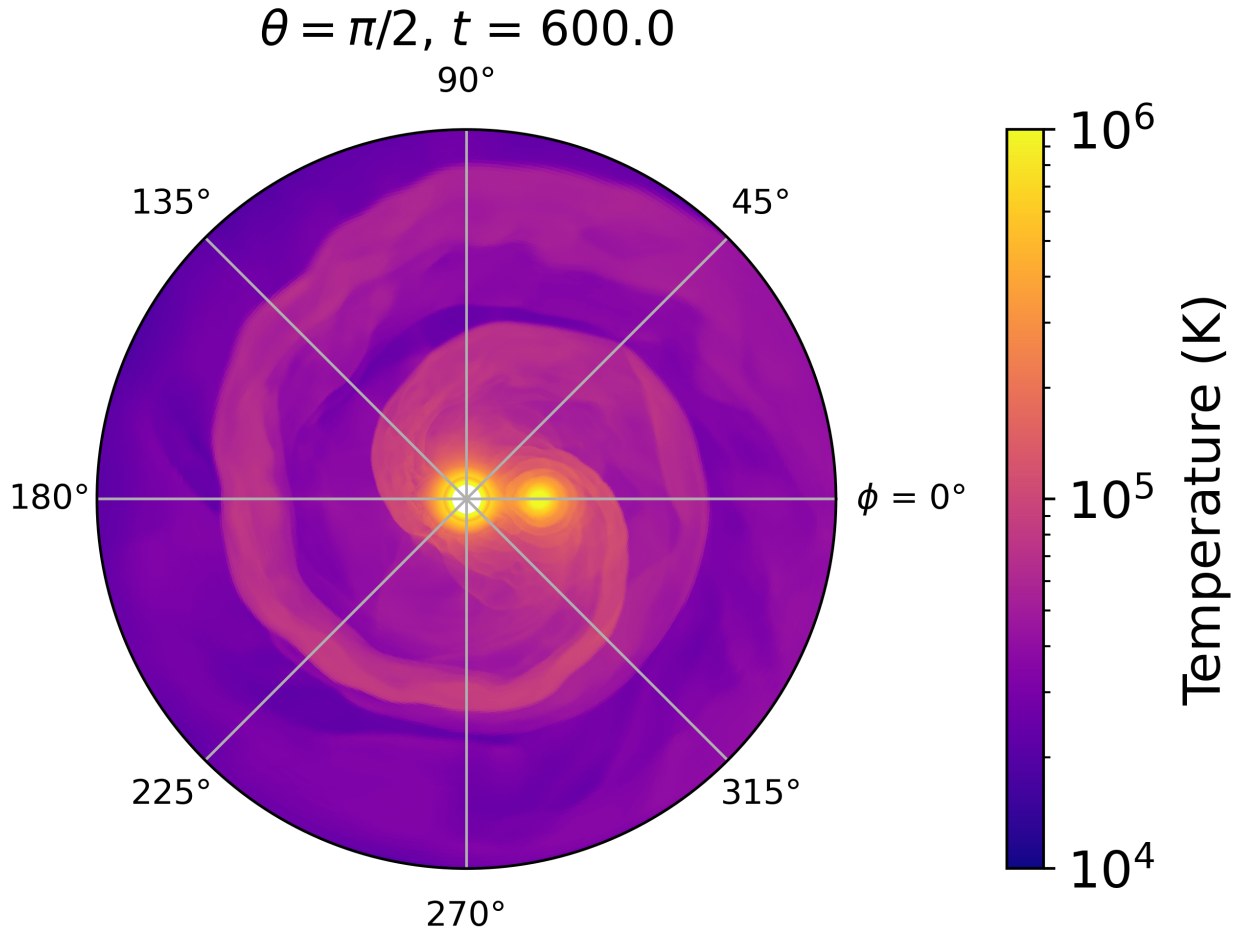

**Figure 13.** An equatorial temperature map of the gas, with the full radial extent of the domain shown, for the $q = 0.5$, intermediate heat simulation q_0.5_mid_heat. To convert from code units to physical units, we assume $M_1 = 10\ M_\odot$, $M_2 = 5\ M_\odot$, and an orbital period of 10 days.

For our $q = 0.5$ intermediate heating simulation q_0.5_mid_heat, with binary parameters of $M_1 = 10\ M_\odot$, $M_2 = 5\ M_\odot$, and an orbital period of 3 days (Example A), the approximate steady-state mass transfer rate in physical units would be 6.9e-2 $M_\odot\ \mathrm{yr}^{-1}$. Once MT in the binary has ramped up, typical values of $\tau$ at different angles are $(\theta = \pi/2, \phi = 0, \tau \gtrsim 10^7)$, $(\theta = \pi/2, \phi = \pi/2, \tau \gtrsim 10^4)$, and $(\theta \approx 0, \phi = 0, \tau \gtrsim 10^3)$. Values of $t_{\mathrm{phot}}$, in units of the orbital timescale of the binary, are $\gtrsim 10^4, \gtrsim 50, \gtrsim 1$ for the same angles. Our assumption of inefficient cooling breaks down closest to the pole, which is unsurprising given the low density in this region outside the donor star. We note that photons likely diffuse out from the disk and circumbinary outflow in the vertical direction, rather than the radial direction, so $t_{\mathrm{phot}}$ may be overestimated. See Appendix A.2 for more discussion of radiative cooling.

If we instead use assume a 100 day orbital period (Example C in Table 1), the MT rate for the same simulation would be 2.1e-3 $M_\odot\ \mathrm{yr}^{-1}$. Typical values of $\tau$ at different angles are instead $(\theta = \pi/2, \phi = 0, \tau \gtrsim 10^5)$, $(\theta = \pi/2, \phi = \pi/2, \tau \gtrsim 10^2)$, and $(\theta \approx 0, \phi = 0, \tau \gtrsim 10)$. Values of $t_{\mathrm{phot}}$, in units of the orbital timescale of the binary, are $\gtrsim 100, \gtrsim 0.1, \gtrsim 0.01$ for the same angles. Therefore, at lower MT rates, our assumptions of inefficient cooling start to break down at all angles.

These results somewhat justify our assumption of inefficient cooling of the outflow at very high MT rates. We note that the gas cooling time is longer than $t_{\mathrm{phot}}$ by a factor of $\sim (P_{\mathrm{rad}} + P_{\mathrm{gas}})/P_{\mathrm{rad}}$, which is large if the outflow is gas-pressure dominated. Additionally, the opacity could be somewhat larger than electron scattering due to free–free and bound–free absorption. However, the neglect of cooling will not be a good approximation at lower MT rates, where the photon diffusion time becomes smaller (scaling approximately with $\dot{M}$).

### *A.2. Observational Appearance*

This work has predicted the presence of an accretion disk as well as dense circumbinary outflows, originating mainly at L2. We characterize the observational signature of the binary system by estimating the photon trapping radius, beyond which photons can free-stream and escape the gas without scattering. Therefore, we calculate where the optical depth equals $c/v$, assuming that the density follows an $r^{-2}$ dependence (Section 3.4), so that

$$\frac{c}{v_{r,\mathrm{inertial}}} = \tau = \int_{r_{\mathrm{trap}}}^{\infty} \kappa\rho dr = \int_{r_{\mathrm{trap}}}^{\infty} \kappa\bar{\rho}_0 \frac{a^2}{r^2} dr \tag{A4}$$

where $\bar{\rho}_0$ is the angle-averaged density at $r = a$, and $a = 1$ in code units. $v_{r,\mathrm{inertial}}$ is defined in Section 3.4, and in the equation above, we use a typical value of $0.15 \times v_{\mathrm{orb}}$. Note that this integral extends to $r = \infty$ and can be evaluated analytically, whereas the estimates of $\tau$ in Appendix A.1 were performed numerically out to the outer boundary of the domain.

We note that the above equation for the photon trapping radius applies even when the outflow only occupies a fraction of the solid angle (as we are using the angle-averaged density), and $r_{\mathrm{trap}}$ only depends on the total mass-loss rate and the radial velocity of the outflow.

The value of $r_{\mathrm{trap}}$ can be calculated for different binary parameters. For $M_1 = 10\ M_\odot$, $M_2 = 5\ M_\odot$, and $a = 0.1$ au (Example A in Table 1), $r_{\mathrm{trap}} \approx (5\text{–}6) \times a$. For $a = 0.22$ au (Example B in Table 1), $r_{\mathrm{trap}} \approx a$. Note that $6a$ is outside the computational domain of our simulations, so we have assumed that typical values of $v_{r,\mathrm{inertial}}$ are valid outside the domain as well to calculate $r_{\mathrm{trap}}$. In Example A, photons would be trapped far out into the circumbinary outflow, whereas in Example B, an observer would see photons emitted from near the accretion disk.

We estimate the radiative flux $F_{\mathrm{rad}}$ and luminosity $L_{\mathrm{rad}}$ emitted at $r = r_{\mathrm{trap}}$, as well as the color temperature $T_{\mathrm{rad}}$, through the following equations:

$$F_{\mathrm{rad}} = -\frac{4c}{3\kappa\rho} a_{\mathrm{rad}} T^3 \frac{dT}{dr} \tag{A5a}$$

$$L_{\mathrm{rad}} = 4\pi r_{\mathrm{trap}}^2 F_{\mathrm{rad}} \tag{A5b}$$

$$T_{\mathrm{rad}} = \left(\frac{F_{\mathrm{rad}}}{\sigma_{\mathrm{SB}}}\right)^{1/4}, \tag{A5c}$$

where $\sigma_{\mathrm{SB}}$ is the Stefan–Boltzmann constant. We find $L_{\mathrm{rad}} \approx 10^{39}\ \mathrm{erg\ s^{-1}}$ for both Example A and B, and $T_{\mathrm{rad}} \approx 10{,}000$ K for Example A and $\approx 20{,}000$ K for Example B. These differences in $T_{\mathrm{rad}}$ and the relative location of $r_{\mathrm{trap}}$ suggest a range of observational signatures associated with the MT we simulate, which will be investigated further in subsequent work. The simplistic analysis above has averaged over angle and neglected any viewing angle effects (but should be accurate as order-of-magnitude estimates), and we defer more sophisticated analysis to future work.

Future work should also perform radiation-hydrodynamic simulations of rapid MT, instead of using pure hydrodynamics as in this work. W. Lu et al. (2023) predicted that ultraviolet photons from a fast disk wind, launched from the inner accretion disk, will be reprocessed into infrared radiation by the circumbinary outflow. Simulations incorporating both accretion and radiation may be able to test this picture.

### *A.3. Calculation of Outflow Opening Angle*

Here we calculate the average opening angle of outflowing gas $\theta_{\rm outflow,av}$, i.e., how closely concentrated to the equatorial plane the outflow is. For each value of $\phi$, we find the angle $\theta'$ at which the density falls by a factor $f_{\rm out}$ relative to the density in the equatorial plane

$$\rho(r > 2,\, \theta',\, \phi) < f_{\rm out} \times \quad \rho(r > 2,\, \theta = \pi/2,\, \phi). \qquad \text{(A6)}$$

Only cells with $r > 2$ are included to avoid areas near the donor and accretor. We perform this calculation for $f_{\rm out} = 1/e$ and $f_{\rm out} = 1/e^2$, representing one or two $e$-foldings. The outflow opening angle is then $\theta_{\rm outflow} = \frac{\pi}{2} - \theta'$, because $\theta'$ is measured from the pole. Finally, we take the average of $\theta_{\rm outflow}$ over $\phi$ to calculate $\theta_{\rm outflow,av}$.

The outflow angle versus time for several simulations is plotted in the bottom panel of Figure 6, for $f_{\rm out} = 1/e$. In general, we find that the angle for $f_{\rm out} = 1/e$ is approximately half that of $f_{\rm out} = 1/e^2$, as expected for an exponential drop-off in density.

### *A.4. Properties of Disk*

The accretion disk around $M_2$ is not the main focus of this work and is not as well resolved as simulations of some other works. However, D. Toyouchi et al. (2024) showed that mass loss from the accretion disk is concentrated near the Lagrange points and is not from a fast wind near the accretor, so resolving the inner disk may not be necessary. This work therefore focuses on the disk outflows and their global quantities.

We find that $v^2$ and $c_s^2$, the velocity and sound speed squared, are comparable in the disk. At several $M_2$-centered spherical shells, i.e., $|\boldsymbol{r} - a\hat{x}| =$ constant, we calculate the density-weighted ratio of $v^2$ and $c_s^2$, i.e.,

$$\left\langle \frac{v^2}{c_s^2} \right\rangle \equiv \frac{\langle \rho v^2 / c_s^2 \rangle}{\langle \rho \rangle} \qquad \text{(A7)}$$

where the $\langle\rangle$ denotes averaging over a spherical shell. Note that this equation uses velocities in the corotating frame, because we wish to use velocities relative to $M_2$.

Figure 14 plots $\langle \frac{v^2}{c_s^2} \rangle$ at different radii around $M_2$, with all radii in units of $a$. Once the simulation reaches a quasi-steady state, the value of $\langle \frac{v^2}{c_s^2} \rangle$ is larger than unity in the inner disk, indicating it is mostly centrifugally supported but with a substantial contribution from pressure. In the outer disk, the value of $\langle \frac{v^2}{c_s^2} \rangle$ is less than unity, indicating pressure support plays a major role, and the disk is very thick.

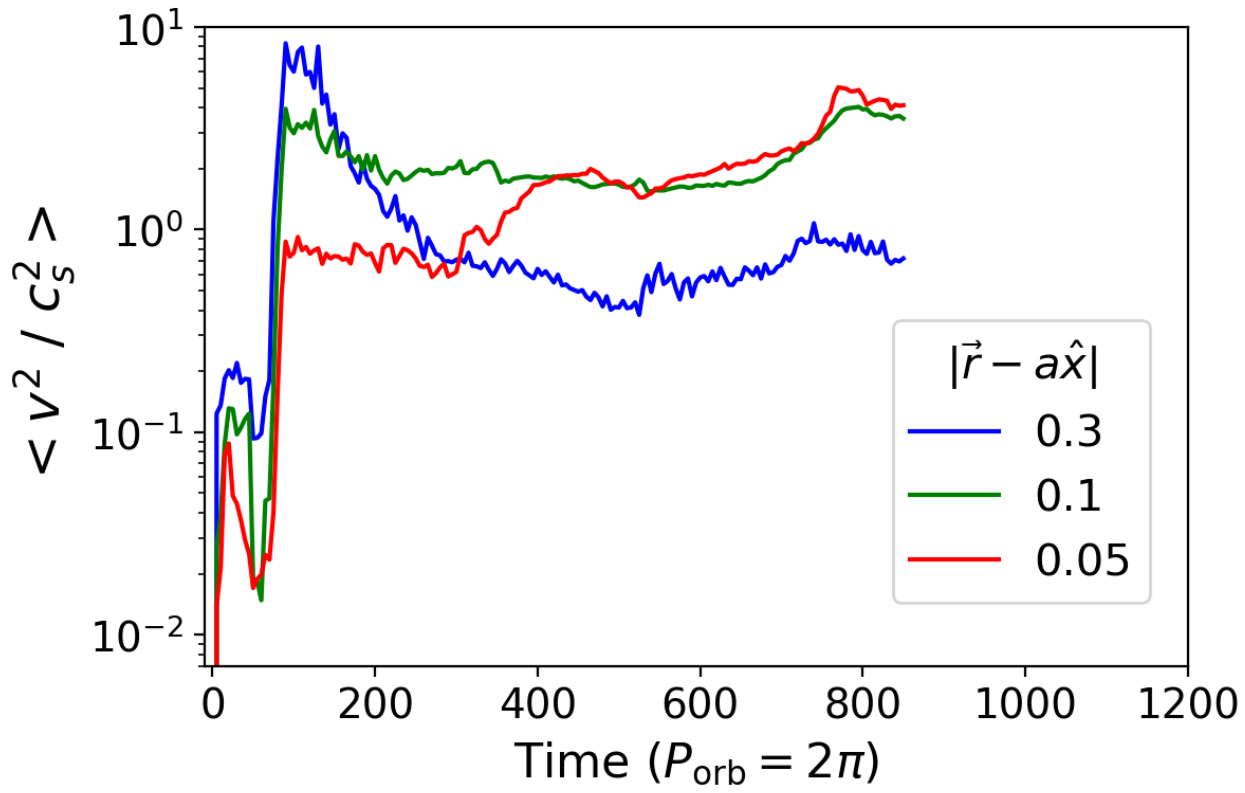

**Figure 14.** The ratio of the velocity squared to the sound speed squared (see Equation (A7)) within the accretion disk, plotted vs. simulation time. This ratio is averaged at several different spheres centered about $M_2$ (which is located at $\boldsymbol{r} = a\hat{x}$). The simulation shown is the same as in Figure 13 above.

## Appendix B
## Outflow at Different Radii

In Section 3.3, we presented our calculation of AM fluxes and $h_{\rm loss}$. To determine the spatial dependence of $h_{\rm loss}$ on distance from the binary, we calculate $h_{\rm loss}$ for several different surfaces centered about the COM and compare the value to $h_{\rm L2}$. Figure 15 plots $h_{\rm loss}/h_{\rm L2}$ as a function of time at several radii for our q_0.5_mid_heat simulation. The largest area with $|\boldsymbol{r} - \boldsymbol{r}_{\rm com}| = r_{\rm max}$ is the maximum COM-centered sphere that will fit on the grid, corresponding to $r_{\rm max} \approx 4.65$ for this mass ratio. The values of $h_{\rm loss}/h_{\rm L2}$ are time-smoothed for each curve, using a rolling average of window-length $t = 50$. The values of $h_{\rm loss}$ are nearly independent of radius and converge to a value near 93% the specific AM of the L2 point. In our other plots involving $h_{\rm loss}$, we quote the value over the surface $|\boldsymbol{r} - \boldsymbol{r}_{\rm com}| = r_{\rm max}$.

Thus, for this mass ratio, escaping mass carries a specific AM similar to that of the L2 point, which is approximately 3 times larger than $h_{M_2}$ for this mass ratio. This approximately supports models where the specific AM of lost material is that of L2.

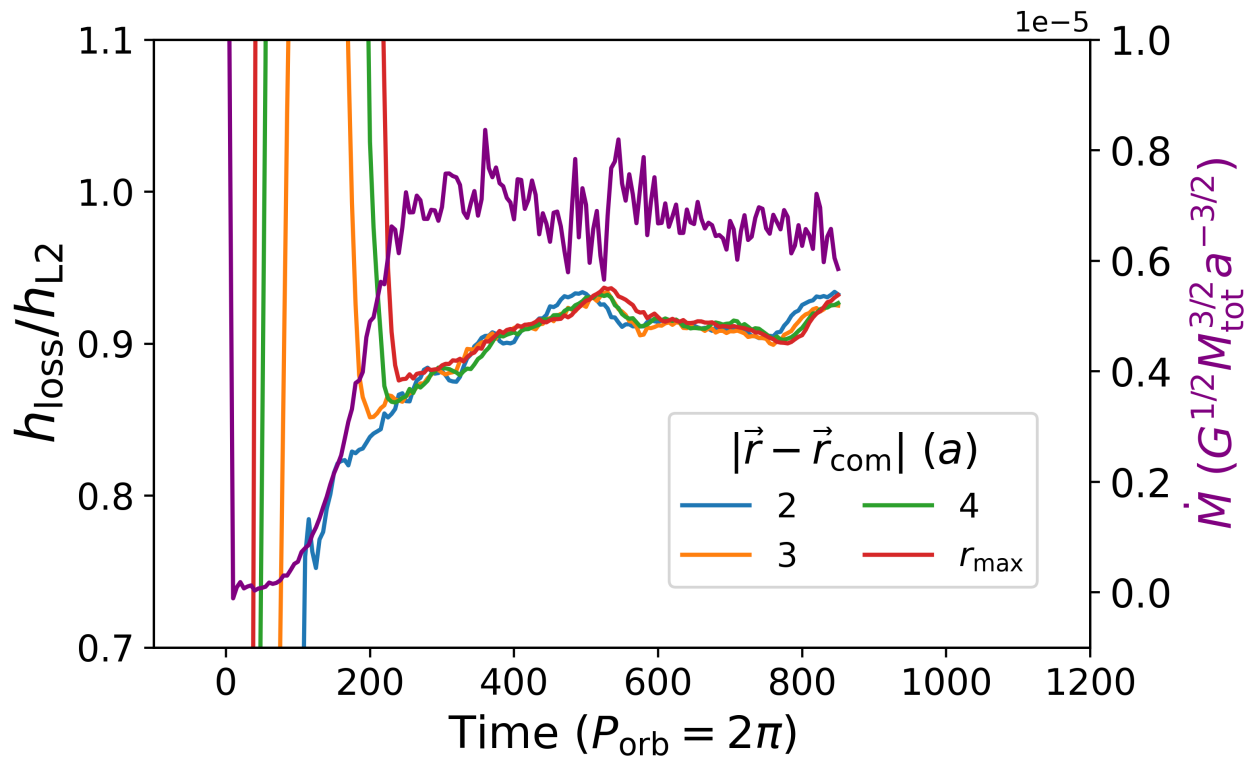

**Figure 15.** Left axis: the specific AM of outflowing material $h_{\rm loss}$, in units of the specific AM of the L2 point $h_{\rm L2}$, is plotted vs. time, for the $q = 0.5$, intermediate heat simulation q_0.5_mid_heat. The value of $h_{\rm loss}$ is calculated for four different spherical shells centered about the binary's center of mass, and smoothed over time with a moving average. Right axis: the mass transfer rate of the donor star. For both quantities, the system reaches a steady state around $t \approx 400$.

## Appendix C
## Convergence Tests

Figure 16 shows convergence testing results for the value of $h_{\rm loss}/h_{\rm L2}$. Included are two 3D simulations of low/high spatial resolution ($\delta r/r = 0.015, 0.01$), and two 2D simulations of high/very high spatial resolution ($\delta r/r = 0.01, 0.005$). Resolution in $\theta$ and $\phi$ is chosen to maintain roughly cubic cells in all cases. The 2D simulations are performed in polar coordinates in the equatorial plane, and reach a quasi-steady state at approximately the same time as the 3D simulations.

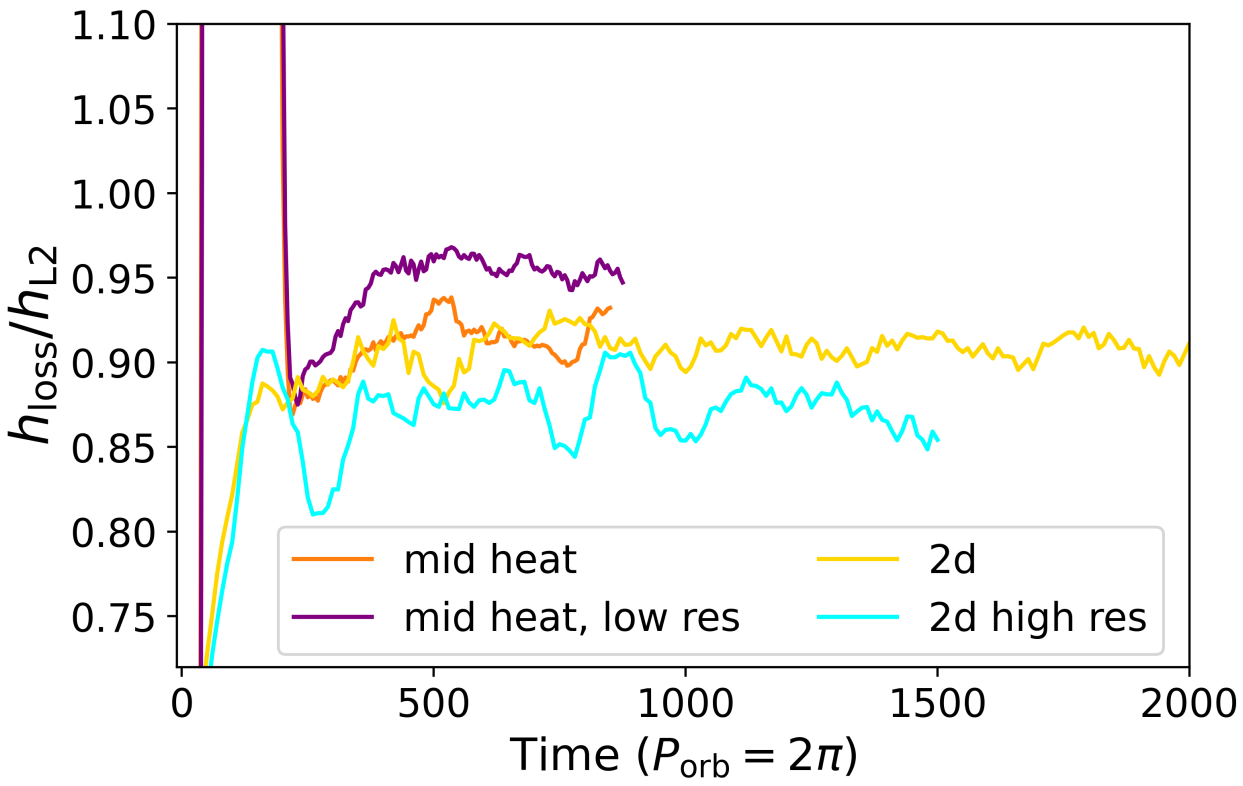

**Figure 16.** Similar to Figure 11, but now showing $h_{\rm loss}/h_{\rm L2}$ for simulations of the same mass ratio $q = 0.5$ but with different spatial resolution. The two 3D simulations, q_0.5_mid_heat and q_0.5_low_res, include intermediate heating and have nearly identical setup, except one has a lower resolution. One 2D simulation has the same resolution as the higher-resolution 3D simulation, q_0.5_mid_heat, whereas the other has approximately double the spatial resolution.

There appears to be a trend of slightly lower values of $h_{\rm loss}/h_{\rm L2}$ with increasing spatial resolution. Future work should investigate this issue in more detail. See also Figure 17, which includes convergence testing for $h_{\rm loss}$ and also outflow velocities, but only for simulations conducted in 3D. See also discussion at end of Section 4.1, which discusses potential pitfalls involving low values of $q$ and insufficient spatial resolution.

### C.1. Effect of Envelope Heating Rate

We investigate the dependence of our results on the envelope heating rate, which drives the expansion of the donor star and sets the mass transfer rate. Figure 17 shows that $\bar{v}'_{r,\rm inertial}$ and $h_{\rm loss}/h_{\rm L2}$ are only weakly affected by different heating rates. See Table 1 for details on the varied luminosity injected and donor expansion timescales. Although the simulations reach a quasi-steady state on different timescales, their steady-state values of $\bar{v}'_{r,\rm inertial}$ and $h_{\rm loss}/h_{\rm L2}$ are similar, with curves reaching similar values. We also include a simulation of lower spatial resolution, such that $\delta r/r = 0.015$, as opposed to our fiducial simulations of $\delta r/r = 0.01$. Otherwise, the lower-resolution simulation is nearly identical in setup to the default intermediate heating simulation. The lower-resolution simulation gives steady-state values that are moderately different from those of the corresponding higher-resolution simulation.

### C.2. Effect of Softening Potential

Figure 18 shows the effect of changing the Plummer softening length $\epsilon$ (Equation (5)) on $h_{\rm loss}/h_{\rm L2}$, for 2D simulations of the equatorial plane. The value of $h_{\rm loss}/h_{\rm L2}$ is similar for 3D versus 2D simulations of similar parameters and resolution, so our results regarding $\epsilon$ can likely be extended to 3D. $h_{\rm loss}/h_{\rm L2}$ slightly decreases with decreasing $\epsilon$, but the steady-state value is similar for $\epsilon = 0.04, 0.05$, and 0.075 and has nearly converged, whereas the value is moderately higher for $\epsilon = 0.1$ (all in units of $a$). Our 3D simulations use $\epsilon = 0.05$. We also tested $\epsilon = 0.025$ and experienced numerical issues such as very high values of the Mach number. Additional work will be needed to understand the effects of small $\epsilon$, which can arise physically from compact object accretors.

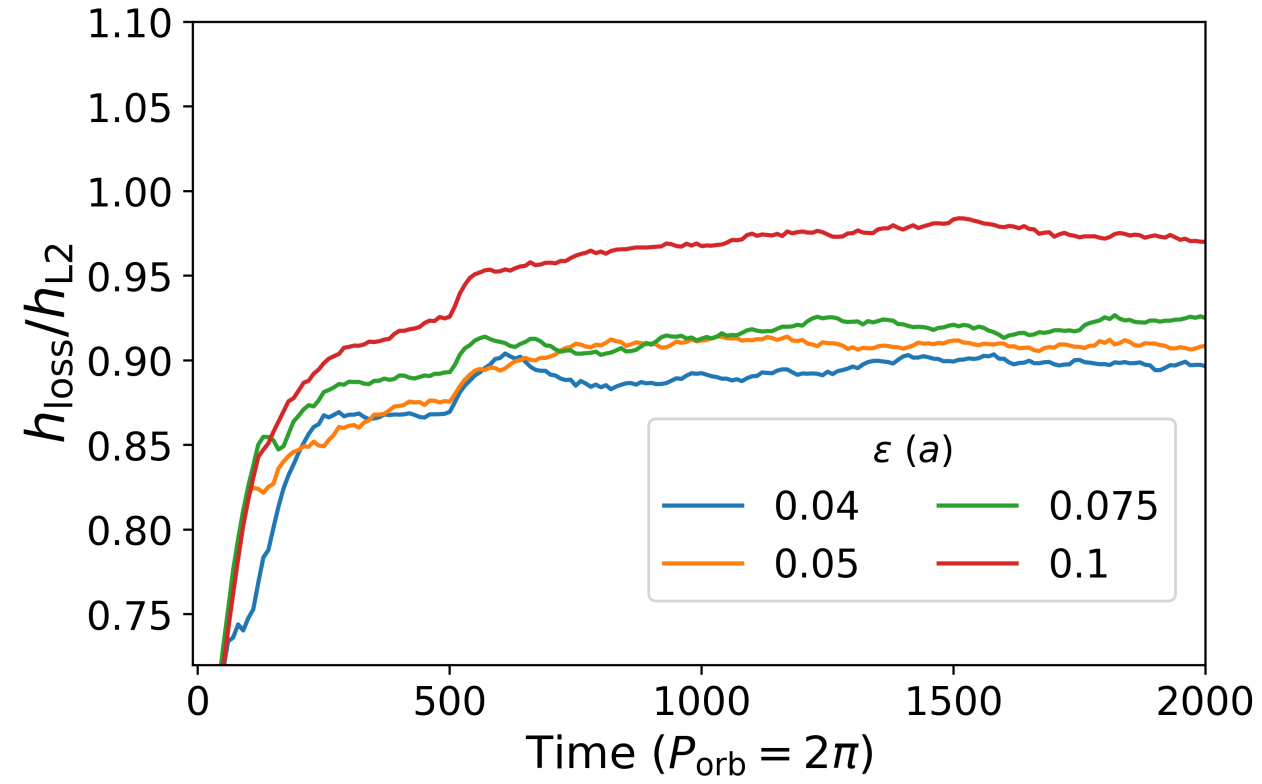

**Figure 18.** Similar to Figures 11 and 16, but now showing $h_{\rm loss}/h_{\rm L2}$ for 2D simulations of mass ratio $q = 0.5$ but varied softening potential $\epsilon$. Our 3D simulations use $\epsilon = 0.05$.

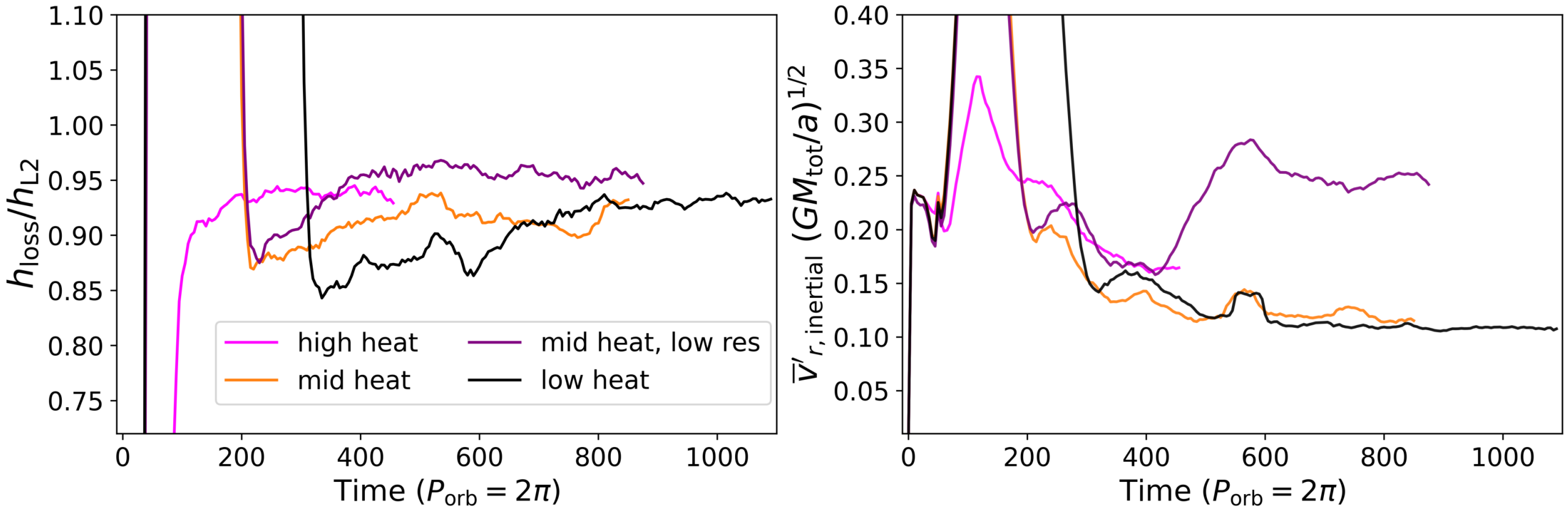

**Figure 17.** A comparison of specific AM and radial velocity of outflowing gas, for simulations of $q = 0.5$ but varying envelope heating (q_0.5_low_heat, q_0.5_mid_heat, and q_0.5_high_heat) or varying spatial resolution (q_0.5_low_res). The low-resolution simulation is otherwise nearly identical in setup to the intermediate heating simulation, q_0.5_mid_heat. The left panel is constructed the same as in Figure 11, and the right panel is constructed the same as in Figure 12, so that the velocity is in units of $\sqrt{GM_{\rm tot}/a}$.

## ORCID iDs

Peter Scherbak https://orcid.org/0000-0003-4221-9097
Wenbin Lu https://orcid.org/0000-0002-1568-7461
Jim Fuller https://orcid.org/0000-0002-4544-0750

## References

Bavera, S. S., Fragos, T., Qin, Y., et al. 2020, A&A, 635, A97
Bavera, S. S., Fragos, T., Zevin, M., et al. 2021, A&A, 647, A153
Blundell, K. M., Mioduszewski, A. J., Muxlow, T. W. B., Podsiadlowski, P., & Rupen, M. P. 2001, ApJL, 562, L79
Bobrick, A., Davies, M. B., & Church, R. P. 2017, MNRAS, 467, 3556
Booth, R. A., Mohamed, S., & Podsiadlowski, P. 2016, MNRAS, 457, 822
Cehula, J., & Pejcha, O. 2023, MNRAS, 524, 471
Chen, Z., Frank, A., Blackman, E. G., Nordhaus, J., & Carroll-Nellenback, J. 2017, MNRAS, 468, 4465
Cherepashchuk, A. M., Postnov, K. A., & Belinski, A. A. 2018, MNRAS, 479, 4844
Church, R. P., Dischler, J., Davies, M. B., et al. 2009, MNRAS, 395, 1127
Clark, P., Maguire, K., Inserra, C., et al. 2020, MNRAS, 492, 2208
Dewi, J. D. M., Pols, O. R., Savonije, G. J., & van den Heuvel, E. P. J. 2002, MNRAS, 331, 1027
Dorozsmai, A., & Toonen, S. 2024, MNRAS, 530, 3706
Ercolino, A., Jin, H., Langer, N., & Dessart, L. 2025, A&A, 696, A103
Fuller, J. 2017, MNRAS, 470, 1642
Gallegos-Garcia, M., Berry, C. P. L., Marchant, P., & Kalogera, V. 2021, ApJ, 922, 110
Gallegos-Garcia, M., Jacquemin-Ide, J., & Kalogera, V. 2024, ApJ, 973, 168
Gottlieb, S., & Shu, C. W. 1998, MaCom, 67, 73
Harris, C. R., Millman, K. J., van der Walt, S. J., et al. 2020, Natur, 585, 357
Hjellming, M. S., & Webbink, R. F. 1987, ApJ, 318, 794
Huang, S. S. 1963, ApJ, 138, 471
Hubová, D., & Pejcha, O. 2019, MNRAS, 489, 891
Hunter, J. D. 2007, CSE, 9, 90
Ivanova, N., Justham, S., Chen, X., et al. 2013, A&ARv, 21, 59
Ivanova, N., Kundu, S., & Pourmand, A. 2024, ApJ, 971, 64
Kadam, K., Motl, P. M., Marcello, D. C., Frank, J., & Clayton, G. C. 2018, MNRAS, 481, 3683
Klencki, J., Istrate, A., Nelemans, G., & Pols, O. 2022, A&A, 662, A56
Klencki, J., Nelemans, G., Istrate, A. G., & Chruslinska, M. 2021, A&A, 645, A54
Klencki, J., Podsiadlowski, P., Langer, N., et al. 2025, arXiv:2505.08860
Kolb, U., & Ritter, H. 1990, A&A, 236, 385
Korol, V., Hallakoun, N., Toonen, S., & Karnesis, N. 2022, MNRAS, 511, 5936
Lajoie, C.-P., & Sills, A. 2010a, ApJ, 726, 67
Lajoie, C.-P., & Sills, A. 2010b, ApJ, 726, 66
Leahy, D. A., & Leahy, J. C. 2015, ComAC, 2, 4
LIGO Scientific CollaborationKAGRA Collaboration, Abbott, R., et al. 2023, PhRvX, 13, 041039
Lu, W. 2025, Tidal Equilibrium of a Star in Roche Potential v1, Zenodo, doi:10.5281/zenodo.15499473
Lu, W., Fuller, J., Quataert, E., & Bonnerot, C. 2023, MNRAS, 519, 1409
Ma, L., & Fuller, J. 2023, ApJ, 952, 53
MacLeod, M., & Loeb, A. 2020a, ApJ, 893, 106
MacLeod, M., & Loeb, A. 2020b, ApJ, 895, 29
MacLeod, M., Ostriker, E. C., & Stone, J. M. 2018a, ApJ, 863, 5
MacLeod, M., Ostriker, E. C., & Stone, J. M. 2018b, ApJ, 868, 136
Marchant, P., Pappas, K. M. W., Gallegos-Garcia, M., et al. 2021, A&A, 650, A107
Mcley, L., & Soker, N. 2014, MNRAS, 445, 2492
Metzger, B. D., & Pejcha, O. 2017, MNRAS, 471, 3200
Mignone, A. 2014, JCoPh, 270, 784
Mignone, A., Bodo, G., Massaglia, S., et al. 2007, ApJS, 170, 228
Mignone, A., Zanni, C., Tzeferacos, P., et al. 2012, ApJS, 198, 7
Mink, S. E., Pols, O. R., & Hilditch, R. W. 2007, A&A, 467, 1181
Mohamed, S., & Podsiadlowski, P. 2007, ASP Conf. Ser. 372, Wind Roche-Lobe Overflow: A New Mass-Transfer Mode for Wide Binaries, ed. R. Napiwotzki & M. R. Burleigh (San Francisco, CA: ASP), 397
Mohamed, S., & Podsiadlowski, P. 2012, BaltA, 21, 88
Nandez, J. L. A., Ivanova, N., & Lombardi, J. C., Jr. 2014, ApJ, 786, 39
Nazarenko, V. V., Glazunova, L. V., & Shakun, L. S. 2005, ARep, 49, 284
Olejak, A., Klencki, J., Xu, X.-T., et al. 2024, A&A, 689, A305
Paczynski, B. 1976, in IAU Symp. 73, Structure and Evolution of Close Binary Systems, ed. P. Eggleton, S. Mitton, & J. Whelan (Cambridge: Cambridge Univ. Press), 75
Pavlovskii, K., Ivanova, N., Belczynski, K., & Van, K. X. 2017, MNRAS, 465, 2092
Pejcha, O., Metzger, B. D., & Tomida, K. 2016a, MNRAS, 455, 4351
Pejcha, O., Metzger, B. D., & Tomida, K. 2016b, MNRAS, 461, 2527
Picco, A., Marchant, P., Sana, H., & Nelemans, G. 2024, A&A, 681, A31
Podsiadlowski, P., Joss, P. C., & Hsu, J. J. L. 1992, ApJ, 391, 246
Postnov, K. A., & Yungelson, L. R. 2014, LRR, 17, 3
Qin, Y., Fragos, T., Meynet, G., et al. 2018, A&A, 616, A28
Quataert, E., & Shiode, J. 2012, MNRAS: Letters, 423, L92
Regös, E., Bailey, V. C., & Mardling, R. 2005, MNRAS, 358, 544
Reichardt, T. A., De Marco, O., Iaconi, R., Tout, C. A., & Price, D. J. 2019, MNRAS, 484, 631
Ryu, T., Sari, R., de Mink, S. E., et al. 2025, arXiv:2505.18255
Sana, H., de Mink, S. E., de Koter, A., et al. 2012, Sci, 337, 444
Schneider, F. R. N., Podsiadlowski, P., & Müller, P. 2021, A&A, 645, A5
Shepard, K., Gies, D. R., Schaefer, G. H., et al. 2024, ApJ, 977, 236
Shu, F. H., Lubow, S. H., & Anderson, L. 1979, ApJ, 229, 223
Taddia, F., Stritzinger, M. D., Sollerman, J., et al. 2013, A&A, 555, A10
Tauris, T. M., Kramer, M., Freire, P. C. C., et al. 2017, ApJ, 846, 170
Tauris, T. M., Langer, N., & Podsiadlowski, P. 2015, MNRAS, 451, 2123
Temmink, K. D., Pols, O. R., Justham, S., Istrate, A. G., & Toonen, S. 2023, A&A, 669, A45
Toro, E. F., Spruce, M., & Speares, W. 1994, ShWav, 4, 25
Toyouchi, D., Hotokezaka, K., Inayoshi, K., & Kuiper, R. 2024, MNRAS, 532, 4826
Tuna, S., & Metzger, B. D. 2023, ApJ, 955, 125
van Son, L. A. C., de Mink, S. E., Renzo, M., et al. 2022, ApJ, 940, 184
Virtanen, P., Gommers, R., Oliphant, T. E., et al. 2020, NatMe, 17, 261
Wellstein, S., Langer, N., & Braun, H. 2001, A&A, 369, 939
Willcox, R., MacLeod, M., Mandel, I., & Hirai, R. 2023, ApJ, 958, 138
Wu, S., & Fuller, J. 2020, ApJ, 906, 3
Wu, S. C., & Fuller, J. 2022, ApJL, 940, L27